%% file: maris_carraro_parisi.tex
\documentclass[onecolumn]{aa4babbage}

\usepackage{natbib}
\usepackage{graphicx}
\usepackage{longtable}

\newcommand{\REPLACE}[2]{{\bf{#2}}}
\newcommand{\SKIP}[1]{}

\newcommand{\lsim}{\,\lower2truept\hbox{${<\atop\hbox{\raise4truept\hbox{$\sim$}}}$}\,}
\newcommand{\gsim}{\,\lower2truept\hbox{${>\atop\hbox{\raise4truept\hbox{$\sim$}}}$}\,}

\newcommand{\AcritSQ}{\mbox{$A^2_\mathrm{crit}$}}
\newcommand{\AcritSQzero}{\mbox{$A^2_\mathrm{crit,0}$}}

\newcommand{\Amax}{\mbox{$A_\mathrm{max}$}}
\newcommand{\PFA}{\mbox{$P_\mathrm{FA}$}}
\newcommand{\PFAzero}{\mbox{$P_\mathrm{FA,0}$}}

\newcommand{\Ac}{\mbox{$A_\mathrm{c}$}}
\newcommand{\As}{\mbox{$A_\mathrm{s}$}}

\newcommand{\Pmin}{\mbox{$P_\mathrm{min}$}}
\newcommand{\Pmax}{\mbox{$P_\mathrm{max}$}}

\newcommand{\Texp}{\mbox{$T_\mathrm{exp}$}}

\newcommand{\Rbck}{\mbox{$R_\mathrm{bck}$}}

\newcommand{\airmass}{\mbox{$\mathrm{Airmass}$}}

\newcommand{\minst}{\mbox{$m_\mathrm{inst}$}}
\newcommand{\Mred}{\mbox{$M$}}

\newcommand{\Sycorax}{Sycorax}
\newcommand{\Stephano}{Stephano}
\newcommand{\Prospero}{Prospero}

\newcommand{\Setebos}{Setebos}
\newcommand{\Trinculo}{Trinculo}
\newcommand{\Uranus}{Uranus}

\def\Aptp{\Delta R_{\mathrm{p2p}}}

\def\AmpR{A_{\mathrm{R}}}
\def\ASy{A_{\mathrm{R}, \mathrm{Sycorax}}}
\def\APr{A_{\mathrm{R}, \mathrm{Prospero}}}
\def\ASe{A_{\mathrm{R}, \mathrm{Setebos}}}

\def\fdump{f_{\mathrm{Dump}}}

\def\micron{\mbox{m$\mu$}}

\begin{document}


\titlerunning{VLT Photometry of 5 Uranian Irregulars}
\authorrunning{M.Maris, G.Carraro, M.G.Parisi}

\title{
Light curves and colours of the faint Uranian irregular satellites 
\Sycorax, \Prospero, \Stephano, \Setebos\ and \Trinculo\ 
\footnote{Based on observations with the ESO {\it Very Large Telescope + FORS2} at the Paranal Observatory, Chile, under program 075.C-0023. }}

  \author{Michele Maris
          \inst{1}
           \and
           Giovanni Carraro
           \inst{2,3}
           \and
           M. Gabriela Parisi
           \inst{4,5,6}
          }

   \offprints{M. Maris}

 \institute{INAF, Osservatorio Astronomico di Trieste, Via
   G.B. Tiepolo 11,  I-34131, Trieste, Italy \\
               \email{maris@oats.inaf.it}
               \and 
             Dipartimento di Astronomia, Universit\`a di Padova,
                  Vicolo Osservatorio 2, I-35122 Padova, Italy\\
                \email{giovanni.carraro@unipd.it}
                 \and
              Andes Prize Fellow,  Universidad de Chile and Yale University
                 \and
              Departamento de Astronom\'ia, Universidad de Chile,
                 Casilla 36-D, Santiago, Chile\\
                  \email{gparisi@das.uchile.cl}
                  \and
               Member of the Centro de Astrofisica,
            Fondo de Investigacion Avanzado en Areas Prioritarias (FONDAP),
             Chile.
              \and
           Member of the Consejo Nacional de Investigaciones Cientificas y Tecnicas (CONICET), Argentina.
}

\date{Received 13 December 2006 / Accepted 17 April 2007 }

\abstract
{
After the work of Gladman~et~al.~(1998), it is now assessed that many
irregular satellites are orbiting around \Uranus.
}
{
Despite many studies have been performed in past years, very few
is know for the light-curves of these objects and inconsistencies
are present between colours derived by different authors.
This situation motivated our effort to improve both the knowledge
of colours and light curves.
}
{
We present and discuss time series observations
of \Sycorax, \Prospero, \Stephano, \Setebos\ and \Trinculo, five
faint irregular satellites of \Uranus, carried out at VLT, ESO Paranal (Chile)
 %
in the nights between 29 and 30 July, 2005 and
25 and 30 November, 2005. 
}
{
We derive light curves for \Sycorax\ 
 and
 \Prospero\
and colours for all of these these bodies.
}
{
For \Sycorax\ we obtain colours 
B-V =$0.839\pm0.014$, V-R = $0.531\pm0.005$ 
and a
light curve which is suggestive of a periodical variation with period
$\approx 3.6$~hours and amplitude $\approx0.067\pm0.004$~mag.
The periods and colours we derive for \Sycorax\ are 
 in agreement with our
previous determination in 1999 using NTT.
We derive also a light-curve for \Prospero\ which suggests an amplitude of
about 0.2~mag and a periodicity of about 4~hours.
However, the sparseness of our data, prevents a more precise characterization
of the light--curves, and we can not determine wether they are 
one--peaked or two--peaked. Hence, these periods and amplitudes 
have to be considered preliminary estimates.
As for \Setebos, \Stephano\ and \Trinculo\ the present data do not allow to derive 
any 
unambiguous periodicity, despite \Setebos\ displays a significant variability
with amplitude about as large as that of \Prospero.
Colours for \Prospero, \Setebos, \Stephano\ and \Trinculo\ are in marginal 
agreement with the literature.
}


\maketitle

\section{Introduction}

In  recent  years  many irregular  satellites  has  been
discovered around \Uranus\ 
\citep{Gladman:etal:1998,Kavelaars:etal:2004,Gladman:etal:2000,Sheppard:Jewitt:Kleyna:2005}.
Irregular satellites  are  those  planetary
satellites on highly elliptic and/or highly inclined (even retrograde)
orbits  with large  semi-major  axis.  These
objects cannot have formed by circumplanetary accretion as the regular
satellites  but  they  are  likely  products  of  captures  from
heliocentric  orbits, probably  in association  with the 
planet formation itself
 \citep{Greenberg:1976,Morrison:Burns:1976,Morrison:etal:1977,jew05}.
It is possible for an object  circling about the Sun to be temporarily
trapped by a planet 
 \citep[][to cite only some]{Heppenheimer:1975,Greenberg:1976,Morrison:Burns:1976}.
But to  turn a temporary capture into a permanent
one  requires a  source of dissipation of orbital energy 
and  that particles  could 
remain  inside the  Hill  sphere long  enough  for the  capture to  be
effective 
\citep{Pollack:etal:1979}.
Otherwise, the trapped object will escape within at most 
few hundred of orbits
\citep{Byl:Ovenden:1975,Heppenheimer:1975,Heppenheimer:Porco:1977,Pollack:etal:1979}.
During the planet formation  epoch several mechanisms  may have  operated,
some of which have the potential to be active even after this early epoch.
They fall mainly  into few categories: 
collisional interactions \citep{Colombo:Franklin:1971},
pull--down capture \citep{Heppenheimer:Porco:1977},
gas drag  \citep{Pollack:etal:1979},
four bodies interactions in the reference frame of the the Sun-Planet system 
either between the captured body and a large
regular satellite of the planet \citep{Tsui:2000},
or between the two components of a binary object leading to
an exchange reaction where 
one of the components of the binary 
is captured and the other is ejected from the system
\citep{Agnor:Hamilton:2006}.

Collisional capture, the so called  break-up process leads to the formation
of  dynamical groupings.   The resulting  fragments of  the progenitor
body after a  break-up will form a population  of irregular satellites
expected  to  have  similar  composition, i.e.,  similar  colours,  and
irregular surfaces.
Large temporal variations in the  brightness of irregular satellites are 
expected from rotating bodies of highly elongated shapes and/or irregular 
surfaces consistent with  a collision fragment origin.

Gas drag is expected to occur in the environment of the protoplanetary 
nebula
\citep{Byl:Ovenden:1975,Hored:1976,Heppenheimer:Porco:1977,Pollack:etal:1979}
and it may origin dynamical families of fragments.
In this case fragments would be produced by the hydrodynamical breaking 
of the intruding body in smaller chunks
in case they exceed the tensile strength of the entering body 
\citep{Pollack:etal:1979}.  
Gravitational attraction of fragments prevents them from escaping,
in general the hydrodynamical pull being not larger than self-gravity, but a small
impact of a $\sim1$~Km size object, likely common
in the nebula environment, would be sufficient to disperse them
without introduce a further fragmentation
\citep{Pollack:etal:1979}.
A specific prediction of this scenario is the production of fragments
with a more regular/round surface than in the break--up process, leading
to light--curve with low amplitude variations
\citep{Pollack:etal:1979}.

On the contrary, if pull--down capture, four bodies interactions
or exchange reactions are the
dominant causes of formation of the irregular satellites, each object would be the 
result of an independent capture event.
In this case, no obvious correlation between dynamical properties, colours and
light--curves would be expected. 

To cast light on these scenarios,
colours and light  curves are very important, since  they would allow one to
discriminate  between   collisional  or  non-collisional   origin  for
irregular satellites.
Theories of irregular satellite  capture have lacked many constraints.
However, the  rapidly-growing number of known  irregular satellites is
now providing  new insights on  the processes of planet  formation and
satellite capture.

A possible origin of that the large obliquity of \Uranus\ is a
giant impact event between the planet and an Earth-sized planetesimal,
occurred  at the  end  of the  epoch  of accretion
 \citep{Slattery:etal:1992,Parisi:Brunini:1997}.
The dynamical and physical properties  of the Uranian
irregular satellites may shed light on their capture mechanism and may
also  witness  the mechanism  leading  to  the  peculiar tilt  of  the
planet's rotation axis
  \citep{Brunini:etal:2002,Parisi:etal:2006}.
 %
For example,
significant  fluctuations  
have been observed in  the  Caliban light--curve
for which
data are consistent with a light--curve with amplitude
$A_{\mathrm{Caliban}} = 0.12 \pm 0.01$~mag and a most probable
period of $\approx 2.7$~hours
as in the
\Sycorax\ light curve, 
 $A_{\mathrm{Sycorax}} = 0.032 \pm 0.008$~mag
with either a period of $\approx 4.1$~hours or 
$\approx 3.7$~hours
\citep{Maris:etal:2001}.
In this regard \citet{Romon:etal:2001} report discrepancies 
in the spectrum they possibly attributed to rotational effects.
All of this seems to support the idea of a collisional scenario.
 However, the existence of a dynamical grouping
\citep{Kavelaars:etal:2004,Sheppard:Jewitt:Kleyna:2005} 
is still debated on the light of the 
colour determination of \citet{Grav:etal:2004b}.
Regrettably,
there seems to be  a lack of consistency between B-V and V-R
colours of different authors for \Sycorax\ and Caliban
\citep{Maris:etal:2001,Rettig:etal:2001,Romon:etal:2001,Grav:etal:2004}.
This may be ascribed to systematic differences in the photometry
and accompanying calibration or, at least for Caliban, to rotational effects.

In an attempt to improve on the situation, in this paper we present and 
discuss new observations of five irregulars of \Uranus,
\Sycorax, \Prospero, \Setebos, \Stephano\ and \Trinculo, obtained
with the ESO Very Large Telescope on Cerro Paranal, Chile. \\
The paper is
organised as follow:
Sect.~\ref{sec:observations} describes observations and data reduction,
light--curves 
are discussed in Sect.~3, while Sect.~4 present the satellites colours.
The conclusions are reported in Sect.~\ref{sec:conclusions}.

\section{Observations and Data Reduction}\label{sec:observations}

We observed the irregular satellites \Sycorax, \Prospero, \Stephano, \Trinculo\ and \Setebos\ 
with the FORS2 camera
\citep{Appenzeller:etal:1998} 
at the focus of VLT {\em Antu} telescope in Paranal, Chile, in 
the two consecutive nights of July~28 and 29, 2005.

We used the standard FORS2 B, V, R, I filters 
 \footnote{See http://www.eso.org/instruments/fors/inst/Filters/ for further details.},
which are very close to the Bessel system.
In particular, the effective wavelength, $\lambda_{\mathrm{eff}}$, and FWHM, $\Delta \lambda$,  for the
filters reported by ESO are 
$\lambda_{\mathrm{eff},\mathrm{B}} = 0.429$~$\micron$, $\Delta \lambda_{\mathrm{B}} = 0.0880$~$\micron$ for B;
$\lambda_{\mathrm{eff},\mathrm{V}} = 0.554$~$\micron$, $\Delta \lambda_{\mathrm{V}} = 0.1115$~$\micron$ for V;
$\lambda_{\mathrm{eff},\mathrm{R}} = 0.655$~$\micron$, $\Delta \lambda_{\mathrm{R}} = 0.1650$~$\micron$ for R;
and
$\lambda_{\mathrm{eff},\mathrm{I}} = 0.768$~$\micron$, $\Delta \lambda_{\mathrm{I}} = 0.1380$~$\micron$ for I.
\Stephano\ and \Trinculo\ were observed in the same frames
so that we observe five objects with just four sequences.
For this reason \Stephano\ R1, R2, $\dots$, V1, V2, $\dots$, frames corresponds
to \Trinculo\ R1, R2, $\dots$, V1, V2, $\dots$, frames.
Each object has been observed in consecutive sequences of frames.
After the end of the sequence for a given object the telescope
switched to the sequence of another object.
Ideally, colours would have to be calculated by combining magnitudes
from frames in the same sequence, in order to limit the rotational
effects.
For each sequence, pointing of the telescope and orientation of the camera 
have been kept fixed.
During the acquisition of each frame the telescope has been tracked at the same 
rate of the target, while
the telescope has been resetted at the default pointing at 
the beginning of each frame in the sequence.
However, given the slow proper motion and the short exposures, the effect
of differential tracking on background stars has been negligible,
background stars does not appear elongated.
Interruptions due to a ToO, mid-night calibrations 
and some minor problem prevent us from
keeping the same sequences in the two nights.
Both nights have been photometric, with average seeing $\approx 1.1$~arcsec.
FORS2 is equipped with a mosaic of two 2k $\times$  4k MIT CCDs 
(pixel size of $15 \times 15$ micron) with a pixel scale,
with the default 2-by-2 binning, of $0^{\prime\prime}.25/pixel$.
The satellite and \cite{Landolt:1992} 
standard stars were centred in CCD $\#1$. Pre-processing of images, which 
includes 
bias and flat field corrections, were done using standard
IRAF\footnote{IRAF is distributed by NOAO, which are operated by AURA under
cooperative agreement with the NSF.} routines. Aperture photometry was then 
extracted 
using the IRAF tool {\tt QPHOT}, both for the standard stars and the satellite,
using a handful of bright field stars to estimate the aperture correction. The 
resulting corrections were small, going from 0.06 to 0.25 in all filters.
A series of $R$ exposures have been taken with the aim to construct a light curve 
and 
search for some periodicity. A few $B$, $V$ and $I$ exposures have been taken as
well to constrain the satellites' colours.
The calibration was derived from  a grand total of 30 standard stars per night in 
the PG0231+051, SA92, PG2331+055, MARK~A, SA111, PG1528+062 and PG1133+099
\cite{Landolt:1992} fields.
The two nights showed identical photometric conditions, and therefore a single 
photometric solution was derived for the whole observing run

 \begin{equation}
 \Mred  =  \minst + \alpha_{\mathrm{m}} - \beta_{\mathrm{m}} \cdot \airmass ;
 \end{equation}

 \noindent
where $\minst = b$, $v$, $r$ or $i$ are the instrumental magnitudes,
$\Mred = B$, $V$, $R$ or $I$ are the reduced magnitudes, $\airmass$
is the airmass and $\alpha_{\mathrm{m}}$ and $\beta_{\mathrm{m}}$
are the calibration coefficients. We obtain
 $\alpha_{\mathrm{m}} = 2.332$, 2.864, 3.112, 2.546
and
 $\beta_{\mathrm{m}} = 0.269$, 0.177, 0.147. 0.150 respectively
for the B, V, R and I bands.
No colour correction have been applied due to the very small
colour term.
A few additional observations of \Prospero\ were acquired on the nights
of November 22 and 25, 2005 in compensation to the ToO.
We reduced the data in the same way as in
the July run, but obtained an independent calibration, being
$\alpha_{\mathrm{m}} = 2.318$, 2.879, 3.007, 2.529 for the B, V, R and
I bands, respectively, which is very similar to the July one.
The list of measures for the five satellites is in Tab.~\ref{tab:mag}.
The table reports the reduced magnitudes, errors and exposure times.
The shortest 
  exposures have been acquired to improve frame centering nevertheless
we report magnitudes from these frames too.
The time column refers to the starting time of each exposure.
No corrections for light travel times have been applied to these data.

 \section{Light Curves}\label{sec:colors:light:curves}

Fig.s~\ref{fig1a},
\ref{fig1b},
\ref{fig1c},
\ref{fig1d}~and~\ref{fig1e}
presents $R$ light curves respectively for \Sycorax, \Prospero,
\Setebos, \Stephano\ and \Trinculo\ for the July 2005 nights.
Each plot is splitted in two subpanels, the left for
the July 29$^\mathrm{th}$ and the right for
July 30$^\mathrm{th}$.
Squares in gray represents the measurements of a common field star
with similar magnitude.
To avoid confusion, error bars for the field star are not reported and
the averaged magnitude is shifted.
For the same reason the few data obtained in November 2005 for \Prospero\ 
listed in the Tab.\ref{tab:mag} are not plotted in Fig.~\ref{fig1b}.

We analyse magnitude fluctuations in the light-curves  trying to assess
first of all whether the detected variations may be ascribed to random
errors, Sect.~\ref{sec:random},
instabilities in the zero point of the calibration,
 Sect.~\ref{sec:zero:point},
or to the effect of unresolved background objects
Sect.~\ref{sec:background:objects}.
In case the variability has been judged significant 
attempt to estimate the amplitude and the period,
Sect.~\ref{sec:periodicity:detection}.
We use both analytical methods and  a Monte Carlo (MC) -- Bootstrap technique.

\subsection{Testing against random fluctuations}\label{sec:random}

The results of a $\chi^2$ test performed on V, R (and eventually B, I)
measures are reported in the first two columns of
Tab.~\ref{tab:random:tests}
(similar results are obtained by a bootstrap on the data).
In this test the hypothesis $H_0$ to be disproved is that the data are
compatible with a constant signal (different from filter to filter)
with random errors as the sole cause of brightness fluctuations.
The table shows that this hypothesis may be discarded for
\Prospero\ and \Sycorax\ with a very high level of confidence.
As for \Stephano\ and \Setebos\ the level is lower but still significant,
whereas for \Trinculo\ the hypothesis can not be discarded at all.
Before considering the case for a periodic variation, the case for
a systematic trend in the brightness is considered, since the
irregular sampling in time prevents the application of robust
de-trendization techniques. As evident from the table,
even this case can be excluded by the present data at a level of
confidence similar to the constant case.

\subsection{Field stars analysis}\label{sec:zero:point}

In order to asses the level of calibration accuracy in an independent manner,
several field stars having magnitudes encompassing those of
the irregular satellites and common to each frame in both nights have
been measured in the same way as the satellites 
\citep[see][for an example of the adopted technique]{Carraro:etal:2006}.
As shown in Fig.s~\ref{fig1a},
\ref{fig1b},
\ref{fig1c},
\ref{fig1d}~and~\ref{fig1e} field stars are characterised by less
wide fluctuations than satellites.
A more quantitative test is obtained taking a set of
variability indicators by which measuring the variability of field
stars and satellites and comparing them.
This is done in Fig.~\ref{fig:variability} where two variability
indicators, the peak-to-peak variation (top) and the RMS (bottom),
for satellites (red spots) and field stars (light blue asterisks)
are plotted against the $R$ magnitude of the objects.
The first important thing to note is the good consistency between 
the two indicators.
While it is evident that the variability of \Sycorax, \Prospero\ and \Setebos\ 
is above the level of variability of field stars, the same is not true for
\Stephano\ and \Trinculo.
In addition, test for the correlation of,
as an example \Sycorax\ or \Prospero\ and the related field stars shows
that they are not significantly correlated.
As an example the correlation coefficient between \Sycorax\ and three field
stars is $C_{\mathrm{Sycorax}, s} = -0.36$, $-0.08$ and $-0.53$.
The probability that this level of correlation can be reached by chance
even if their time series are not correlated are respectively
$27\%$, $78\%$, $15\%$,
while for \Prospero\ $C_{\mathrm{Prospero}, s} = -0.10$, $0.22$ and $-0.12$ with
probabilities respectively of $64\%$, $36\%$ and $59\%$.
In addition, even assuming a correlation between field stars
fluctuations and satellite fluctuations, it would explain only
a small fraction of the satellites variability. 
Indeed, variances of field stars accounts for 
$0.02\% - 6\%$ of the \Sycorax\ variance,
$6\% - 9\%$ of the \Prospero\ variance,
$12\% - 18\%$ of the \Setebos\ variance,
$2\% - 14\%$ of \Stephano\ or \Trinculo\ variances.
In conclusion, field stars variability, connected to calibration instabilities,
can not account for most of the variability of at least \Sycorax, \Prospero\ and 
\Setebos, while as evident from Fig.~\ref{fig:variability}
\Stephano\ and \Trinculo\ would have to be considered more cautiously.

 \subsection{Unresolved background objects}\label{sec:background:objects}

Unresolved objects in background may affect photometry.
A quicklook to the frames
  \footnote{A sequence of the observed frames for \Prospero\ 
  has been already published in \cite{Parisi:etal:2006}.}
shows that in general, the satellites passes far enough
from background objects to allow a proper separation of their
figures. 
It can be considered also the case in which a satellite
crosses the figure of an undetected faint object in the background
producing a fake time variability
of the satellite light-curve. It is quite simple to compute the
upper limit for the magnitude of a background object,
$\Rbck$,
able to produce
the observed variations of magnitude for these satellites.
The result is $\Rbck \le 23.6$~mag, 24.7~mag, 25~mag, 25.5~mag and
25.5~mag, respectively for \Sycorax, \Prospero, \Setebos, \Stephano\ and
\Trinculo.
Those magnitudes are within the detection limit for
\Sycorax\ and \Prospero, near the detection limit for
\Setebos\ and marginally outside the detection limit for \Stephano\
and \Trinculo.
In conclusion at least for \Sycorax\ and \Prospero, we can be confident
that background is not important.

 \subsection{Looking for amplitudes and periodicities}\label{sec:periodicity:detection}

Given the sparseness of our data it is not easy to asses safely 
the shapes, the amplitudes and the periods of our light-curves,
despite at least for \Sycorax, \Prospero\ and probably \Setebos\ 
a significant variability is present.
However, we think important to attempt a recover of such informations,
at least as a step toward planning of more accurate observations.

Constraints on the amplitude for the part of light curves sampled
by our data may be derived from the analysis of the peak-to-peak 
variation. After excluding data with exposure times below 100~sec
we evaluate peak--to--peak variations for R, $\Aptp$.
Denoting with $\AmpR$ the amplitude of the light--curve we may assume
$\AmpR \approx \Aptp/2$.
To cope with the random noise, $\Aptp$ have been evaluated by 
MonteCarlo, simulating the process of $\Aptp$ evaluation assuming
that the random errors of the selected data are normally distributed.
We obtain
$\ASy \ga 0.07 \pm 0.01$~mag,
$\APr \ga 0.27 \pm 0.04$~mag and 
$\ASe \ga 0.31 \pm 0.05$~mag.
Where the $\ga$ symbol is used because the $\AmpR \approx \Aptp/2$ relation
is strictly valid only if the true minima and maxima of light--curves are
sampled, a condition which we are not safe to have fullfilled.
Another order of magnitude estimate of 
the amplitudes is based on the analysis of 
their RMS, $\mathrm{std}(R_t)$. 
It easy to realize that for any periodical light-curve
of amplitude $\AmpR$ then 
$\mathrm{std}(R_t) \approx \fdump \, \AmpR$.
Where $\fdump>0$ is a factor which depends both on the sampling function 
and the shape of signal. Assuming a sinusoidal signal sufficiently well sampled,
$\fdump \approx 1.39  - 1.46$, giving for \Sycorax\ 
$\ASy \approx  0.06$~mag, 
while for \Prospero\ and \Setebos\ 
$\APr \approx \ASe \approx 0.2$~mag.

We then consider the case for a periodical variation in our light--curves
by attempting first to search for the presence of periodicities
in the hypothesys of a sinusoidal time dependence, and
in case of a positive answer, assessing the most likely sinusoidal amplitude.
To cope with the limited amount of data increasing both
the sensitivity to weak variations and the discrimination power against
different periods, we fit the
same sinusoidal dependence on V, R (and eventually B and I)
assuming that colours are not affected by any significant rotation
effect. In short the model to be fitted is

 \begin{equation}\label{eq:model:to:fit} 
 M_f  =  A \cos({2\pi}/{P} (t - \tau) + \phi) + M_{f,0} \; ,
 \end{equation}

 \noindent
where $f=$~R, V (and eventually B and I) indicates the filter, $M_f$ the
measured magnitudes for that filter, $M_{f,0}$ the averaged magnitude for the
filter $f$, $A$ is the amplitudes $\phi$ the phase
and $\tau$ an arbitrary origin in time.
It has to be noted that phase--angle effects are not considered here.
The reason is that no safe dependence of the magnitude on the
phase--angle has been established so far for these bodies.
On the other hand,  the variation of the phase angles
over two consecutive nights is just about 2.4~arcmin.
Consequently, we expect the phase--angle effect to be quite negligible.
As widely discussed in literature the problem of searching for periodicities
by fitting a model with a sinusoidal time dependence is equivalent to
the analysis of the periodogram for the given data set
\citep[just to cite some]{Lomb:1976,Scargle:1982,Cumming:etal:1999,Cumming:2004}.
The Lomb and Scargle (hereinafter LS) periodogram  is the most
used version.
In this view, a better formulation of the problem is obtained expressing the model in
the following form:

 \begin{equation}\label{eq:sin:model}
 M_f  =  \Ac \cos \left( \frac{2\pi}{P} (t - \tau) \right ) + \As \sin \left( \omega (t - \tau) \right) + M_{f,0}
 \end{equation}

 \noindent
where $\omega = \frac{2\pi}{P}$, $\Ac$ and $\As$ co-sinusoidal and sinusoidal
amplitudes related to $A$ and $\phi$ by the simple relations
$A = \sqrt{\Ac^2+\As^2}$ and phase $\phi = \arctan(\Ac/\As)$.
The time origin $\tau$ can be arbitrarily fixed, but the canonical choice is
\citep{Lomb:1976,Scargle:1982}

 \begin{equation}\label{eq:tau}
    \tan (2 \omega \tau) = \frac{\sum_{f=B,V,R,I} \sum_{j=1}^{N_f}
\frac{\sin(\omega t_{f,j}) }{\sigma^2_{f,j}}}
                              {\sum_{f=B,V,R,I} \sum_{j=1}^{N_f} \frac{\cos( \omega t_{f,j})}{\sigma^2_{f,j}}} \;.
 \end{equation}

 \noindent
which will be our definition of $\tau$.
Then, the free parameters involved in the minimisation are $P$, $M_{f,0}$ for
$f=$V, R (B,I), $\Ac$ and $\As$ or equivalently, $A$ and $\phi$.
However, being interested to amplitude and not to the phase we marginalise
our statistics over $\phi$.

A set of best possible combinations of $M_{f,0}$ for $f=$V, R (B,I), $\Ac$ and $\As$
for each given $P$ in a suitable range $[\Pmin, \Pmax]$ is obtained by minimising

 \begin{equation}\label{eq:chisq}
 \chi^2(P) = \sum_{f=B,V,R,I} \sum_{j=1}^{N_f}
            \frac{
         \left(\Ac \cos(\frac{2\pi}{P}(t_{f,j} - \tau) ) + \As \sin(\frac{2\pi}{P}(t_{f,j} - \tau) ) + M_{f,0} - m_{f,j}\right)^2}
               {\sigma^2_{f,j}},
 \end{equation}

\noindent
where $m_{f,j}$ are the magnitudes for filter $f$ measured
at the times $t_{f,j}$ with associated errors $\sigma_{f,j}$
with $j=1$, 2, $\dots$, $N_f$ the index of $N_f$ measures
obtained for the filter $f$. The minimisation is carried out analytically,
with the $\tau$ defined in Eq.~(\ref{eq:tau}).
This way the method becomes a generalisation of the floating average periodogram 
\citep{Cumming:etal:1999,Cumming:2004} and reduces to it in the case in which magnitudes
come from a single filter, while in the case of homoscedastic data and null zero points
we return to the classical LS periodogram.\\
The sensitivity to noise as a function of $P$ is not constant and varies with time.
Fig.~\ref{fig:time:window} represents the result for a Montecarlo
designed to asses the sensitivity to noise of the periodogram in wide
range of periods for \Sycorax\ (upper frame) and \Prospero\ (lower frame).
The Montecarlo code generates simulated time series
assuming the same time
sampling of data, the same errors, normal distribution of errors and
time independent expectations, and then computes the corresponding periodogram.
Dots in the figure represents the realization of such periodograms,
the full-line represents the Fourier transform of the time window.
The sparseness of that causes a strong aliasing with diurnal periodicities.
The plot is dominated by the prominent 24~hrs diurnal peak, followed by the 
12~hrs, 8~hrs, 6~hrs, 4~hrs, 2~hrs peaks of decreasing amplitude. 
It is evident that above a period of 20~hrs the sensitivity of the periodogram to
random errors is rapidly increasing. Then, our data set is best suited to detect
periods below 20~hrs, or better, due to the presence of the 12~hrs peak, periods 
below 10~hrs.

\noindent
Significant periodicities for $P < 10$~hrs
are likely  present in the data if at least
for one $P$ in the range, the squared amplitude
$A^2(P) = \Ac^2(P)+\As^2(P)$ obtained by minimising
Eq.~(\ref{eq:chisq}), exceeds a critical value \AcritSQ,
which is fixed by determining the false alarm probability
for the given $[\Pmin, \Pmax]$,
$\PFA(A^2>\AcritSQ)$
 \footnote{The $\PFA(A>\AcritSQ)$
 is the probability that random errors are responsible for
 the occurrence of a peak of squared amplitude $A^2 > \AcritSQ$
 in the interval $[\Pmin, \Pmax]$ of interest.}.
This interval has been sampled uniformly with a step size $\Delta P = 0.025$~hours
(the results do not depend much on the choice of the step size)
and the \PFA\ as a function of $\AcritSQ$ has been assessed.
In determining the \PFA\ we exploit the fact that we want to calculate
this probability for values of $\AcritSQ$ for which \PFA\ is small.
In this case,

 \begin{equation}
   \PFA(A^2>\AcritSQ) \approx \PFAzero \exp(-\AcritSQ/\AcritSQzero),
 \end{equation}

\noindent
which is good for $\PFA < 0.3$,  and with the
parameters $\PFAzero$ and $\AcritSQzero$
determined from Montecarlo simulations.
The last two columns of Tab.~\ref{tab:random:tests} report the results
of this generalised version of the LS periodogram.
Again, for \Sycorax\ and \Prospero\ a quite significant periodical signal
is detected. For \Setebos\ the detection is marginal, while for \Stephano\ 
and \Trinculo\ no detection can be claimed at all.
There are many reasons for the lack of detections of periodicity despite
random noise cannot account for the variability.
Among them a
lack in sensitivity, the fact that the period is outside the optimal search
window, and that the 
light-curve cannot be described as a sinusoid.
In conclusion, it is evident that a significant variability is present
in most of these data sets and that
for \Sycorax\ and \Prospero\ 
the July 2005 data suggests the possibility to construct
a periodical light curve.

\noindent
A period can be considered a good candidate if
i.) $\chi^2(P)$ has a local minimum;
ii.) the amplitude $A(P)$ of the associated sinusoid is significantly above the noise;
iii.) the period $P$ is not affected in a significant manner by aliasing with
the sampling window.
Of course one has to consider the fact that one period may be preferred to another
one just by chance. Random fluctuations may lead to a different
selection of the best fit period. We assess this problem by generating random
realizations of time series with expectation given by the measured values of
each sample and $\sigma$ fixed by their Gaussian errors.
For each generation the period producing the minimum $\chi^2(P)$ has been determined.
The probability of selection of each $P$, $P_\mathrm{sel}(P)$, has been then derived.
We then add a criterion iv) that a period $P$ is selected as most likely
if it has the maximum $P_\mathrm{sel}(P)$.
Tab.~\ref{tab:sin:fit} reports the results of the fit. Note the
difference in $\chi^2$ between the best fit with a sinusoid and the
$\chi^2$ in Tab.~\ref{tab:random:tests}.\\

Before looking at the results, 
it has to be stressed that light--curves
can be either single peaked or double peaked.
In the 
 \REPLACE{first}
 {
 second
 } 
case the rotation period will be twice the 
light -- curve period.
We have not enough data to discriminate between these two cases,
therefore the rotational periods of the observed objects could be twice 
the light--curve periods.

\begin{figure}
\centering
 \begin{tabular}{cc}
 \vspace{-2cm}
 \includegraphics[angle=-90,width=8cm]{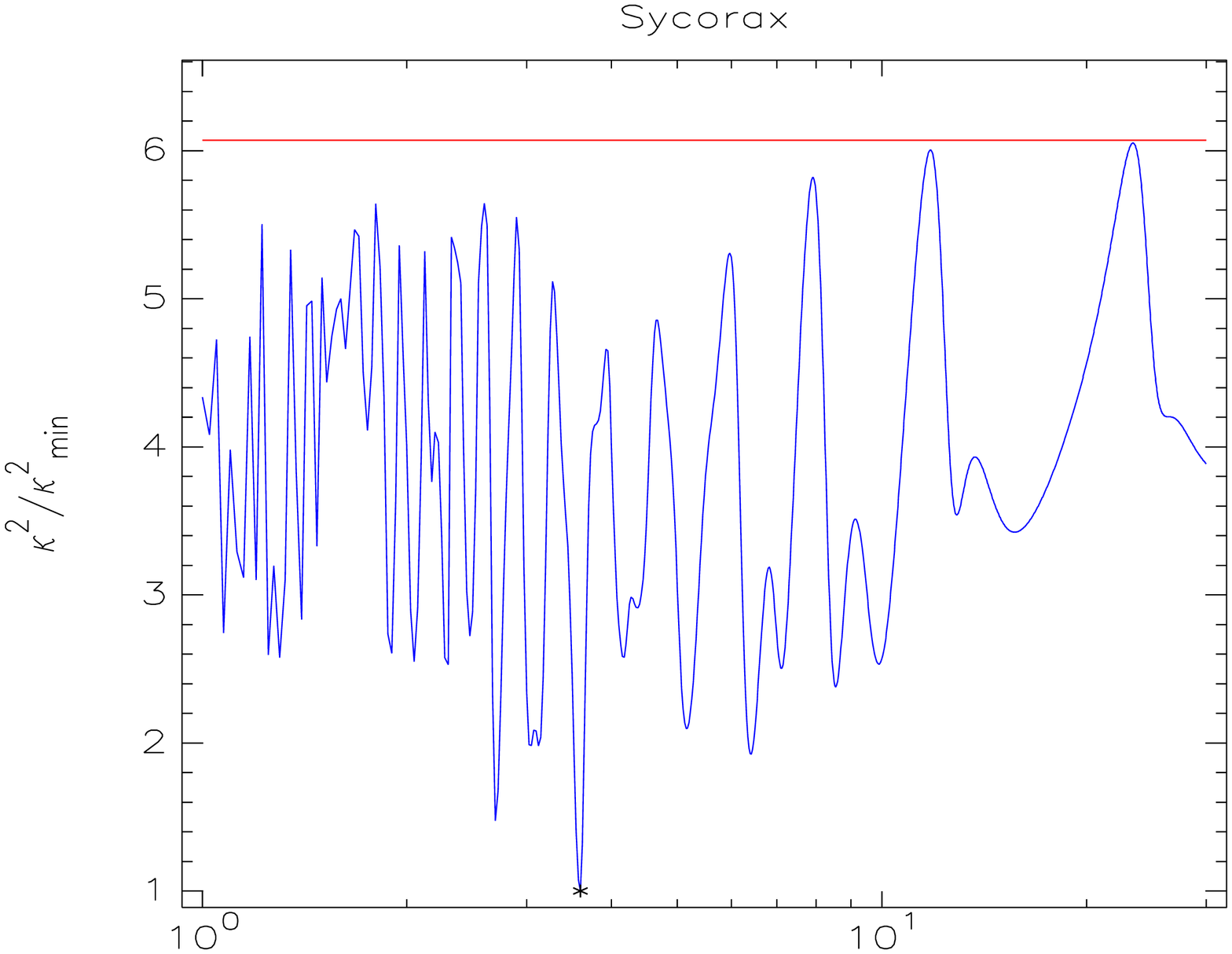} &
 \includegraphics[angle=-90,width=8cm]{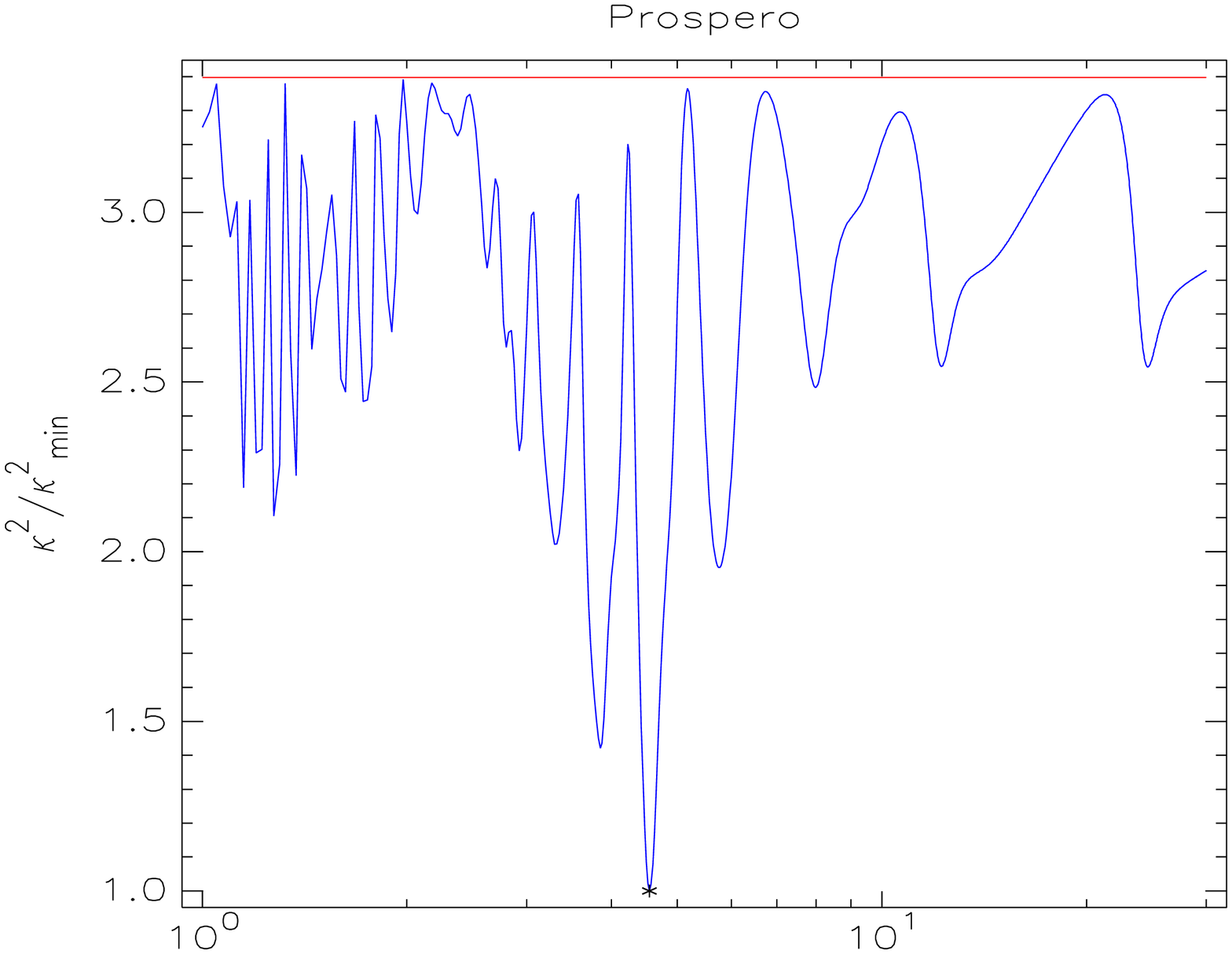} \\
 &\\
 &\\
 &\\
 \includegraphics[angle=-90,width=8cm]{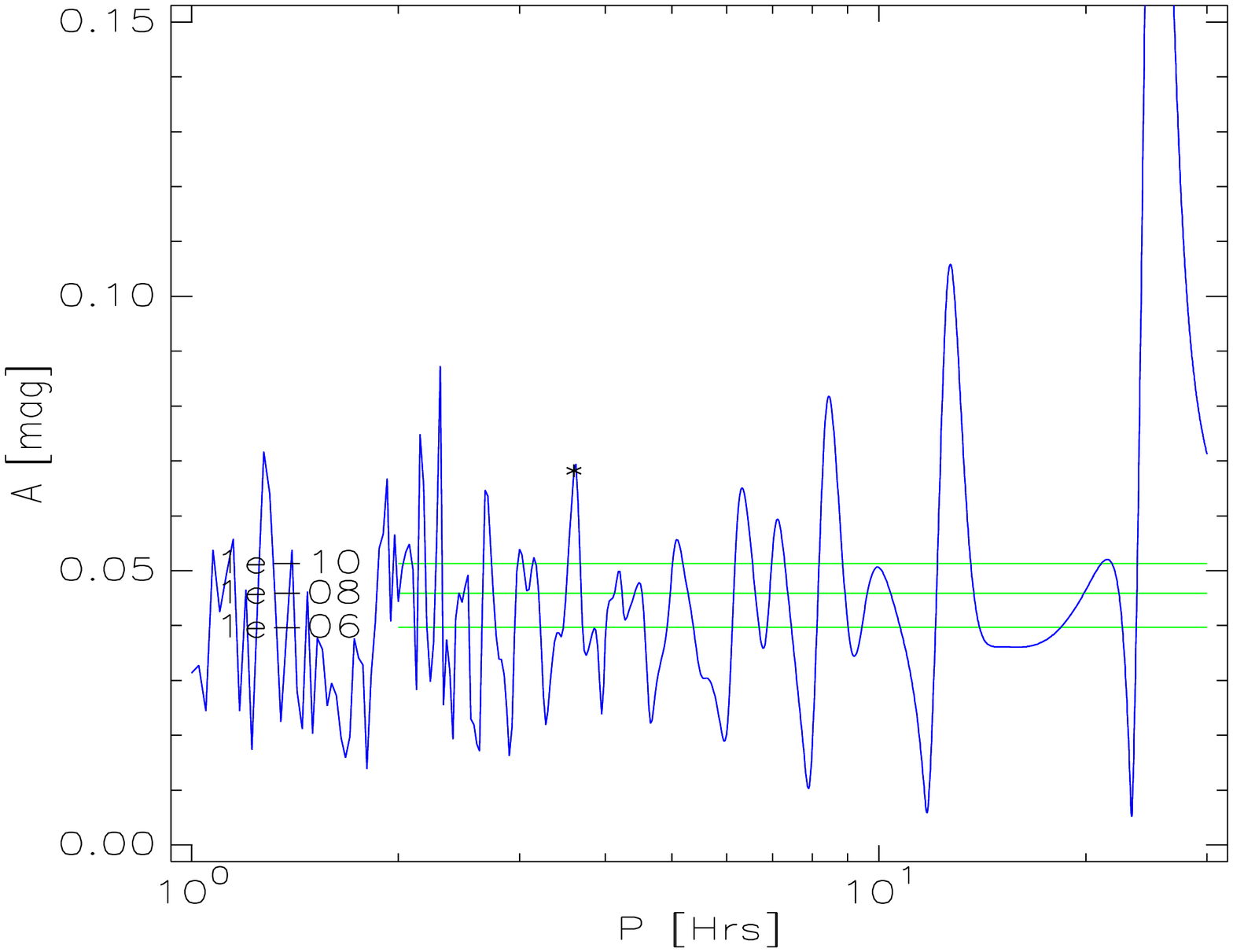} &
 \includegraphics[angle=-90,width=8cm]{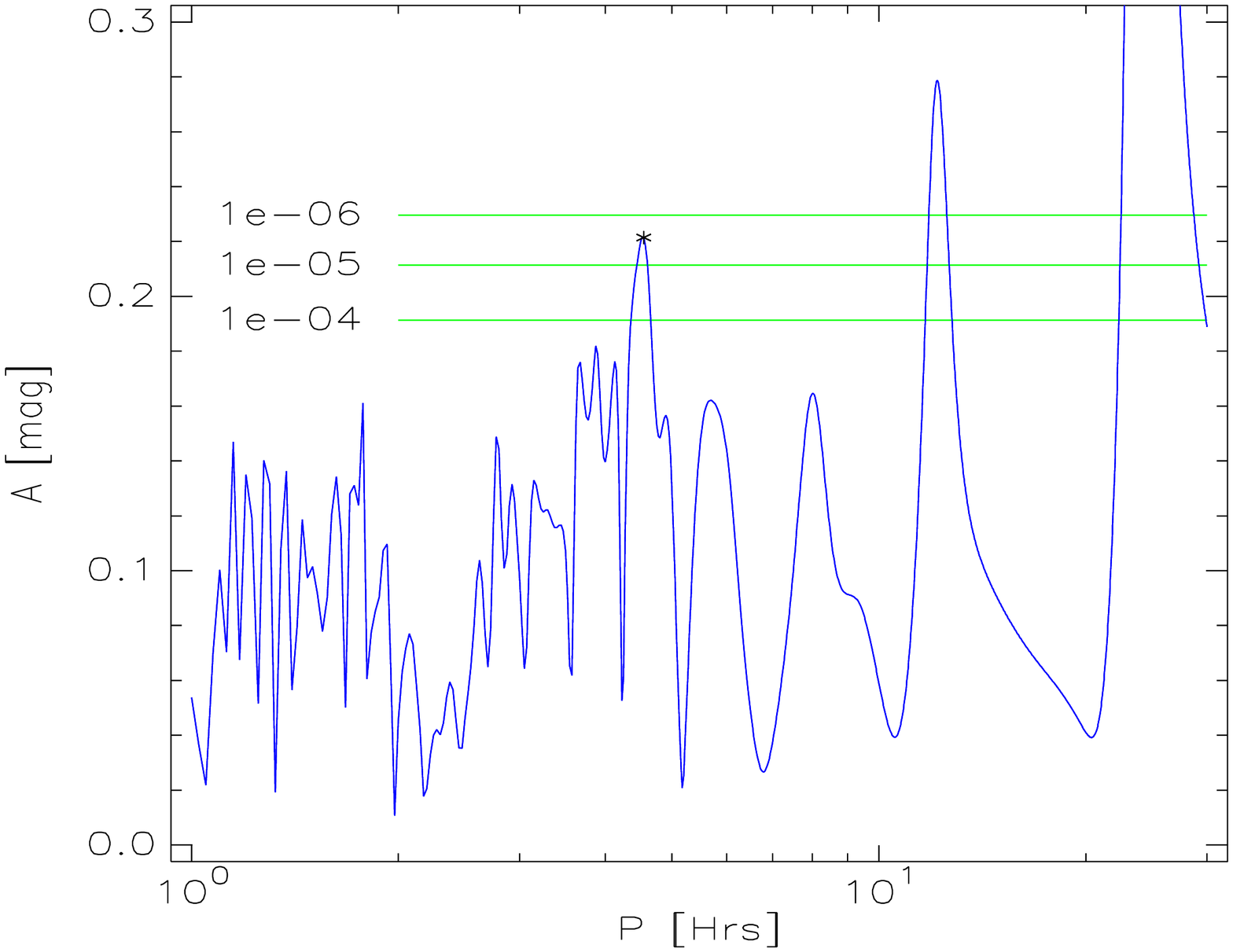} \\
 \includegraphics[angle=-90,width=8cm]{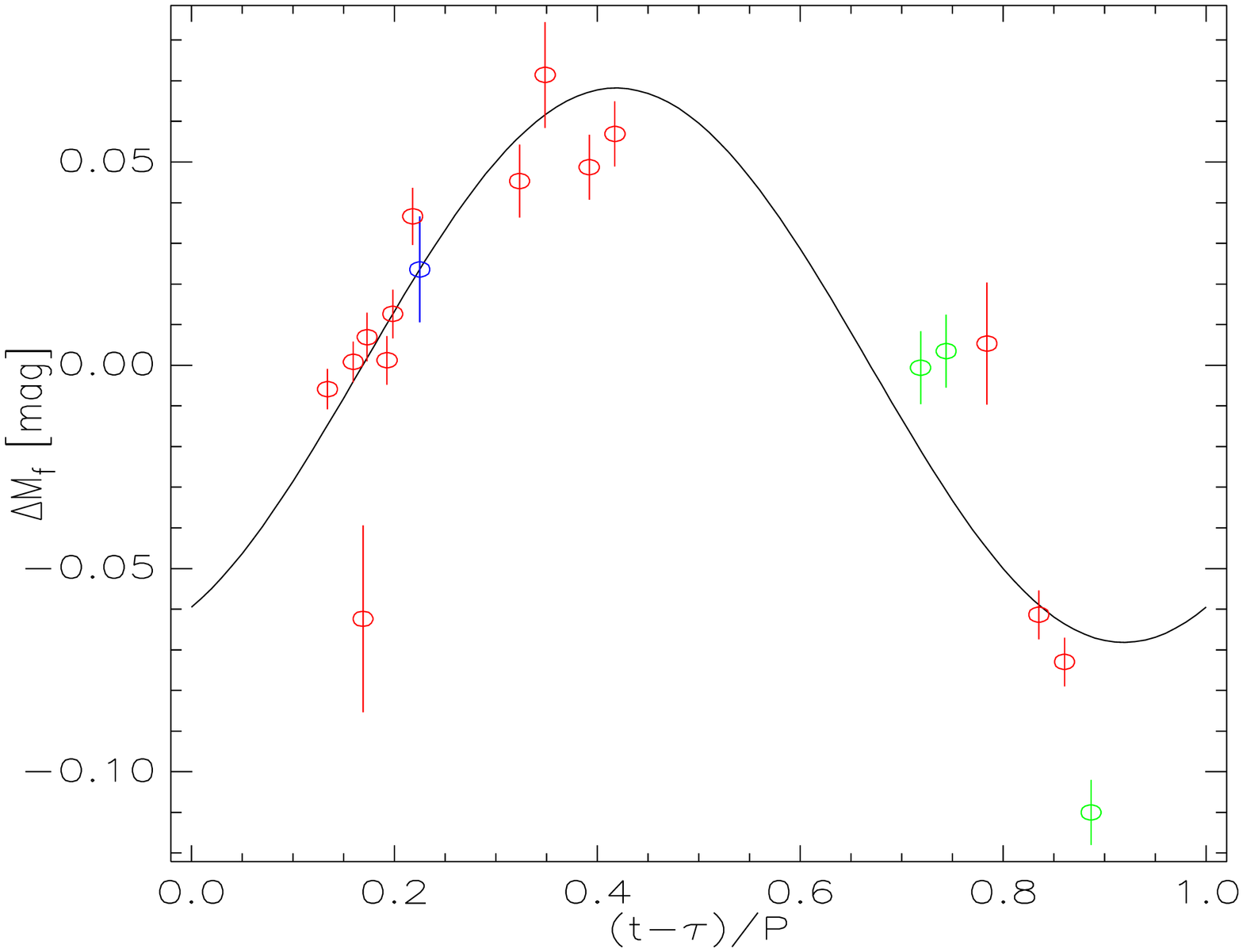} &
 \includegraphics[angle=-90,width=8cm]{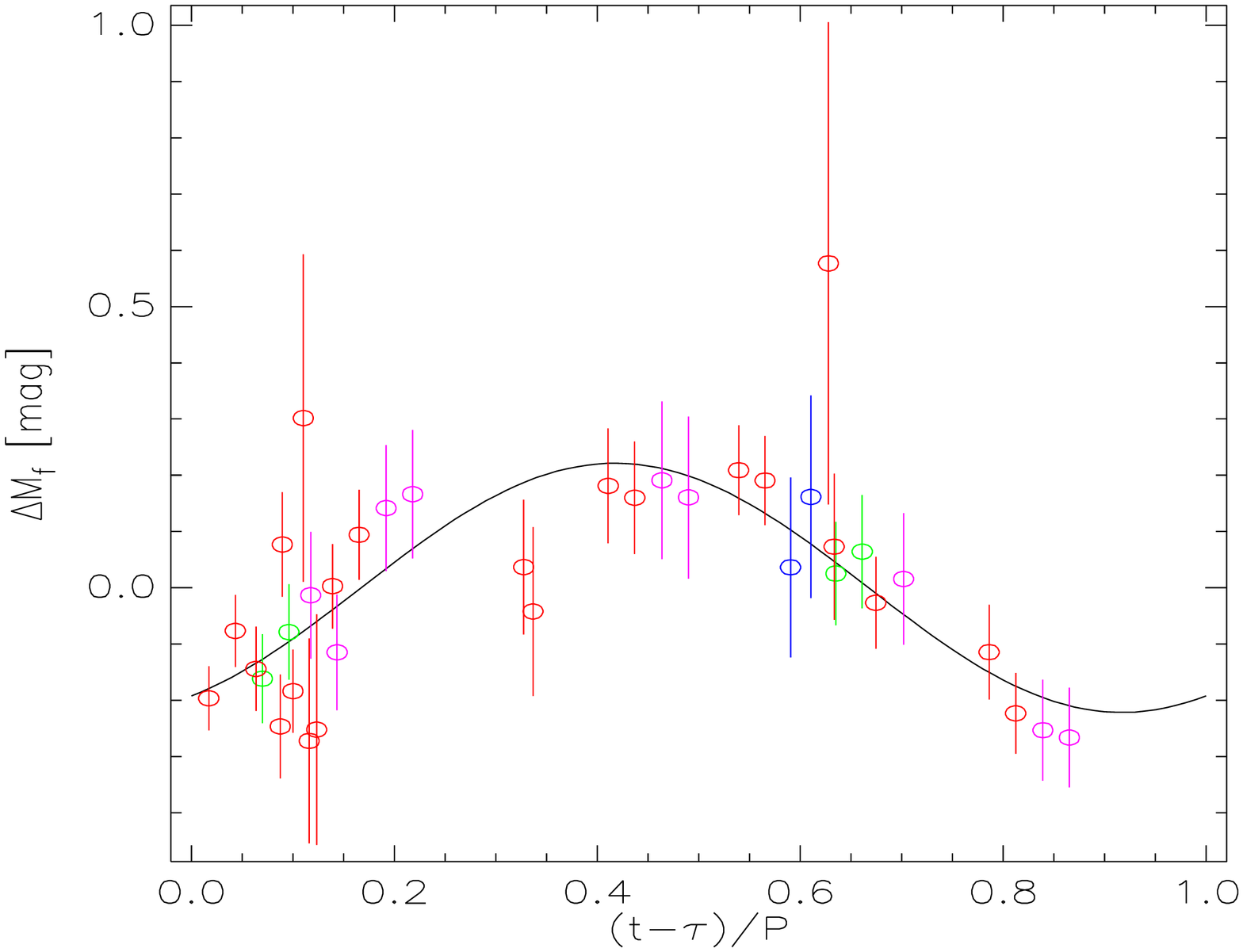} \\
 \end{tabular}
\caption{
LS test of periodicity for \Sycorax\ (left) and \Prospero\ (right)
light-curve as a function of the period $P$. Upper
frames display $\chi^2/\chi^2_{\mathrm{min}}$, the red
line is the $\chi^2/\chi^2_{\mathrm{min}}$ for the case of a constant
signal.
The central frames are the estimated amplitude, a group of green lines
denotes the amplitudes for $\PFA= 10^{-4}$, $10^{-5}$, $10^{-6}$, $10^{-8}$ and $10^{-10}$.
In all the frames a black $*$ marks the best fit point.
Note that the diurnal $12$~hrs and $24$~hrs periods
are excluded by the $\chi^2/\chi^2_{\mathrm{min}}$ despite the largest
amplitude.
The lower frames represents the overlap of the best reconstructed
sinusoid with data ordered
on a folded time scale, $(t-\tau)/P$,
colours are blue for B, green for V, red for R and magenta for I.
  \label{fig:ls}
}
\end{figure}

\noindent
{\bf \Sycorax} \\
Fig.~\ref{fig:ls} on the left is the periodogram for \Sycorax.
The $\chi^2$ suggests that the three periods $P\approx3.6$, 3.1 and 2.8~hr
are favoured with a very high level of confidence
($\PFA < 10^{-8}$).
Bootstrap shows that the first period is the preferred one in about
96.6\%
of the simulations. The third period is chosen in less than 3.4\% of
the cases, the second one instead is chosen in less than 0.01\% of
the cases.
For the best fit case we obtain also
$B_0 = 21.676 \pm 0.013 $,
$V_0 = 20.849 \pm 0.005 $,
$R_0 = 20.276 \pm 0.003$
which are compatible with the weighted
averages discussed in the next section
and are fairly independent of $P$.
The same holds for the ``derotated colours''
$B_0 - V_0 = 0.828 \pm 0.014 $,
$V_0 - R_0 = 0.573 \pm 0.006 $.

\noindent
{\bf \Prospero} \\
Fig.~\ref{fig:ls}~ shows on the right the periodogram for \Prospero.
Given a safe dependence between the phase angle and magnitude
is not established for these bodies, we just used the data taken on July 2005
to evaluate the periodogram.
Four periods are allowed at a $5\sigma$ confidence level (c.l.)
for $P\approx4.6$, 3.8, 5.7 and and 3.3~hours, respectively.
Bootstrap shows that the first peak is the preferred one in about 91\%
of the simulations. The second peak is chosen in less than 6\% of
the cases, the other peaks instead are chosen in less than 3\% of
the cases.
Comparing the periodogram with the spectral window it is evident how
the secondary peaks
are close to alias of the diurnal 24~hr and 12~hr periods.
Removing a 24~hr sinusoid from data before to perform the fit
depresses the 24~hr peak, but not the 4.6~hr peak.
On the contrary the removal of the 4.6~hr component strongly depresses
the spectrum in the range $P\approx 3 - 8$~hr.
By fitting separately the first and second night data, and avoiding the
implicit 24~hr periodicity, the preferred period is 4.3~hr.
Another way to filter the diurnal 24~hr periodicity is to shift
in time the lightcurve of the second night to overlap the
lightcurve of the first. Even in this case periods between 3 and 5
hours look favoured.
The spectral window for the data shows a leakage corresponding to
$P\approx4.3$~hours. A secular variation may introduce power at
periods longer than 48~hours which should leak power at
$P\approx4.3$~hours.
The periodogram for simulated data with a linear time dependence
has a peak in
the $P\approx 4.3 - 4.6$~hours region, but in order to have an
amplitude in the periodogram of 0.2~mag a peak-to-peak variation
in the simulated data about seven or eight times larger than the
peak-to-peak variation observed in real data is needed.
Exclusion of B and/or I data, or fitting only the R data does not
change significantly the results.
The same holds if we remove  the three R points with the largest errors.
As a consequence, the data suggests $P\approx4.6$~hr with $A\approx0.21$~mag.
The lower frame of Fig.~\ref{fig:ls} represents the variations of
B, V, R and I magnitudes folded over the best fit sinusoid;
For the best fit case we obtain also
$B_0 = 24.584 \pm 0.123$,
$V_0 = 23.841 \pm 0.053$,
$R_0 = 23.202 \pm 0.020$,
$I_0 = 22.805 \pm 0.043$ which are compatible with the weighted
averages discussed in the next section
and are fairly independent of $P$.
The same holds for the ``derotated colours''
$B_0 - V_0 = 0.743 \pm 0.134 $,
$V_0 - R_0 = 0.639 \pm 0.057 $,
$R_0 - I_0 = 0.397 \pm 0.047 $.

\noindent
{\bf \Setebos, \Stephano\ and \Trinculo}\\
For the other three bodies no strong evidences are found for a
periodicity in the present data. However for \Setebos\ 
the data may be considered suggestive of some
periodicity with $\PFA \approx 0.7\%$,
the preferred period being $P\approx4.38\pm0.05$~hours with
$A \approx 0.189\pm0.038$~mag.
For \Stephano\ and \Trinculo\ best fit periods are
$P_{\mathrm{Stephano}}\approx2$~hours and
$P_{\mathrm{Trinculo}}\approx5.7$~hours
with amplitudes
$A_{\mathrm{Stephano}}\approx0.459$~mag and
$A_{\mathrm{Trinculo}}\approx0.422$~hours
but with \PFA\ of 22\% and 44\%, respectively.

\section{Averaged Magnitudes and Colours}

The derivation of colours would have to take in account the
removal of rotational effects from magnitudes.
Otherways systematic as large as peak--to--peak variations in
the light--curve can be expected.
Lacking a good light--curve we may apply two possible methods are:
hierarchical determination of colours and compare weighted averages of magnitudes.
The hierarchical method is based on the comparison of magnitudes
from consecutive frames in the hypothesys that time differences
are smaller than the light--curve period, so that rotational effects 
can be neglected. An example is the estimation of \Setebos\ V-R by taking
V3 and R15. In case in which one of the frames obtained with a given filter
X is located between two frames of another filter Y, 
the Y magnitude at the epoch of which the X filter has been acquired 
can be derived by simple linear interpolation.
An example is given by the estimate of V-R for \Prospero\ in  Nov 22, 
by interpolating R2 and R3 at the epoch of V1.
Given in this way different estimates of each colours are obtained, 
weighted averages of such estimates are reported.
The second method is based on the hypothesys that the light--curve has
a periodical behaviour, and that it is 
so well sampled that weighting averages of magnitudes will cancell out
the periodical time dependence.
Of course the first method would be affected by larger random errors being 
based on a subset of the data. The second method is more prone to systematics.

Colours derived by using the first method for \Sycorax, \Prospero\ and \Setebos\ 
are presented in Tab.~\ref{tab:itp:col}.
When more independent estimates of the same colour are possible their weighted 
average is taken. 

Colours derived from the second method
are listed in Tab.~\ref{tab:aver:mag}.
The weighted averages of magnitudes for all of the satellites
are listed in Tab.~\ref{tab:mag}.
In both cases, tables report just random errors and not the systematic calibration
error, which amount to $0.018-0.022$~mag, depending on the filter.

It is evident how the results of the two methods are similar. Hence, 
we present for conciseness the results of the second one as
commonly reported in literature. 

\noindent
{\bf \Sycorax}\\
We obtain $B-V = 0.89 \pm 0.01$ and $V-R = 0.53 \pm 0.01$, which
are both compatible within 1$\sigma$ with the \citet{Maris:etal:2001} determinations
but incompatible with \citet{Grav:etal:2004}.

\noindent
{\bf \Prospero}\\
We have two sets of colour measures for \Prospero,
one from the July and the other from the November run.
We obtain  $B-V = 0.89 \pm 0.13$,  $V-R = 0.63 \pm 0.05$
and $R-I = 0.40 \pm 0.04$ from the July run, and
$B-V = 0.85 \pm 0.13$,  $V-R = 0.59 \pm 0.07$
and $R-I = 0.33 \pm 0.10$. The two set are consistent.
With respect to  \citet{Grav:etal:2004}, we obtain a redder $V-R$,
and a compatible $B-V$.

\noindent
{\bf \Setebos}\\
We obtain  $B-V = 0.74 \pm 0.14$ and $V-R = 0.52 \pm 0.04$.
In this case  \citet{Grav:etal:2004} obtained their data with the Keck II
telescope and DEIMOS, which hosts a rather special filter set.
Our $B-V$ is compatible with their, while as for \Prospero\ our  $V-R$
is redder.\\

\noindent
{\bf \Stephano\ and \Trinculo}\\
For these two extremely faint satellites, we could
only derive the $V-R$ colour, which is $0.73 \pm 0.17$ and
$0.82 \pm 0.43$ for \Stephano\ and \Trinculo, respectively. While
our $V-R$ for \Trinculo\ is in good agreement with  
\citet{Grav:etal:2004},
their $V-R$ for \Stephano\ is very low, and inconsistent with our 
one.

 \section{Summary and Conclusions}\label{sec:conclusions}

In this paper we report accurate photometric B, V, R, I observations
obtained with the VLT telescope in two consecutive
nights in July 2005 of
the Uranian irregular satellites \Sycorax, \Stephano, \Trinculo\ and
\Setebos.
Additional observations of \Prospero\ obtained
in November 22 and 25 of the same year are also reported.

\noindent
From the analysis of the data we conclude that
 %
\Sycorax\ seems to displays a variability of $\approx 0.07$~mag,
apparently larger than our previous result \citep{Maris:etal:2001},
while the period of 3.6~hr is in agreement with our previous 2001 determination,
and the same is true for the colours, so it seems unlikely that the
difference in amplitude can be ascribed to some systematic. 
If true, a possible explanation would be that 
in the two epochs two different parts of the same
light curve have  been sampled. 
But also it has to be noted that larger brightness variations 
have been not reported by other observers in the past years
\citep{Gladman:etal:1998,Rettig:etal:2001,Romon:etal:2001}.
 %
\Prospero\ light--curve exhibits an apparent periodicity 
of 4.6~hr and an amplitude of 0.21~mag. 
The impact of such a sizeable amplitude is throughly discussed in 
\citep{Parisi:etal:2006}.
Colours for \Prospero\ obtained in July and November are in a quite good agreement,
further assessing the goodness of the relative calibration.
 %
\Setebos\ colours are only in marginal agreement with previous studies.
In addition the \Setebos\ light curve displays a significant variability 
but it is not possible to asses a good fit using a simple sinusoidal time
dependence.
Wether this is due to undersampling of the light-curve or to a non-sinusoidal
time dependence can not be decided from these data alone. 
However, assuming a sinusoidal time dependence, our data are suggestive
of a light--curve amplitude of $\approx0.18$~mag with a period 
of $\approx4.4$~hours
which will have to be confirmed or disproved by further observations.
 %
As for \Stephano\ and \Trinculo, the present data do not
allow us to derive any sizeable 
time dependence,
while the colours we derive are in marginal agreement with previous studies
on the subject.

\begin{acknowledgements}
GC research is supported by {\it Fundacion Andes}.
MGP research is supported by FONDAP (Centro de Astrof\'idica, Fondo
de investigac\'ion avanzado en areas prioritarias).
MM acknowledges FONDAP for financial support during a visit to
Universidad the Chile. Part of the work of MM has been also supported by
INAF FFO - {\em Fondo Ricerca Libera} - 2006.
\end{acknowledgements}



\onecolumn

\input maris_carraro_parisi_figures.tex

\clearpage \FIGLIGHTCURVES
\clearpage \FIGVARIABILITY
\clearpage \FIGTIMEWINDOW

\clearpage


\clearpage

\input maris_carraro_parisi_tab1.tex

\onecolumn\clearpage

\input maris_carraro_parisi_tab2.tex

\input maris_carraro_parisi_tab3.tex

\input maris_carraro_parisi_tab4.tex

\input maris_carraro_parisi_tab6.tex

\input maris_carraro_parisi_tab5.tex

\end{document}

%% file: maris_carraro_parisi_figures.tex



\newcommand{\FIGLIGHTCURVES}{
\def\LCSCALE{0.62}
\begin{figure*}
\includegraphics[angle=-90,scale=\LCSCALE]{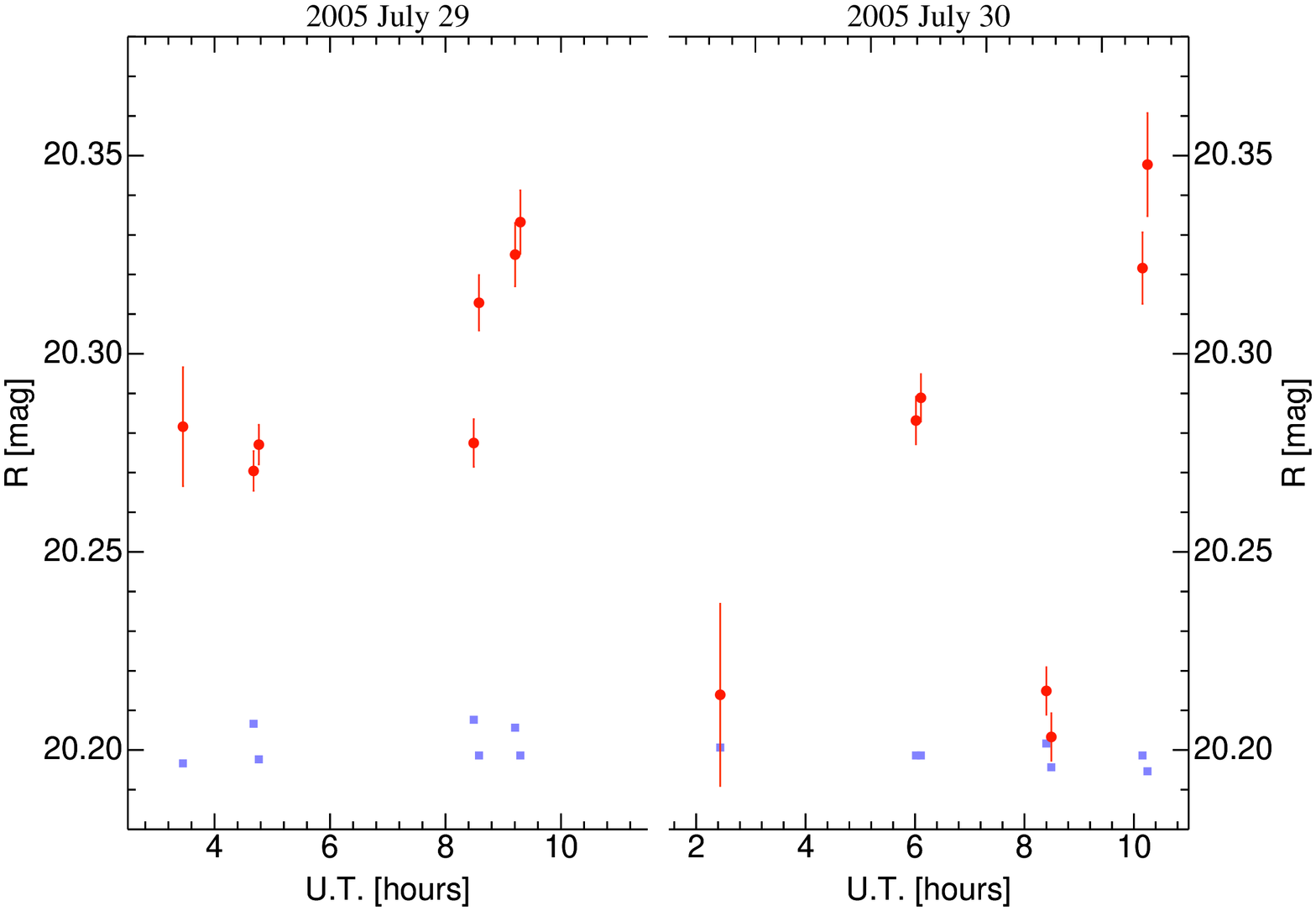}
\caption{
Light curves in $R$ for \Sycorax.
Data are for the nights of \SKIP{2006} 2005, July 29$^\mathrm{th}$
(left) and 30$^\mathrm{th}$ (right).
\SKIP{UT for the second night starts from 24.0 hours instead of 00.0 hours.}
Squares in gray represents measurements of magnitudes for a common field star
of similar magnitude.
To avoid confusion error bars for the field star are not reported and
the averaged magnitude is shifted.
(See the electronic edition of the Journal for a colour version
of this figure).
  \label{fig1a}}
\end{figure*} 

\begin{figure*}
\includegraphics[angle=-90,scale=\LCSCALE]{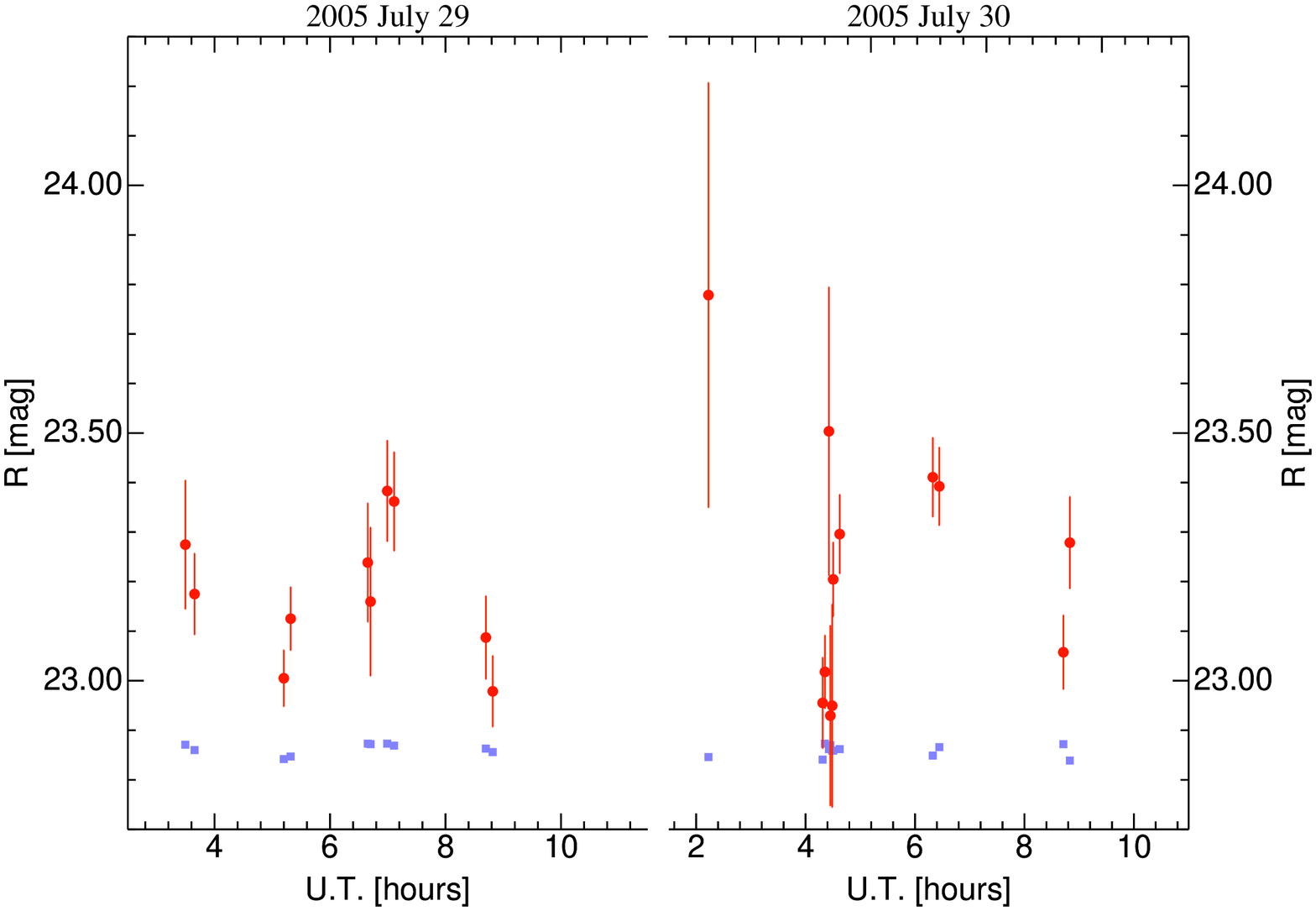}
\caption{
Light curves $R$ for \Prospero.
(See Fig.~\ref{fig1a} for comments).
  \label{fig1b}}
\end{figure*} 

\begin{figure*}
\includegraphics[angle=-90,scale=\LCSCALE]{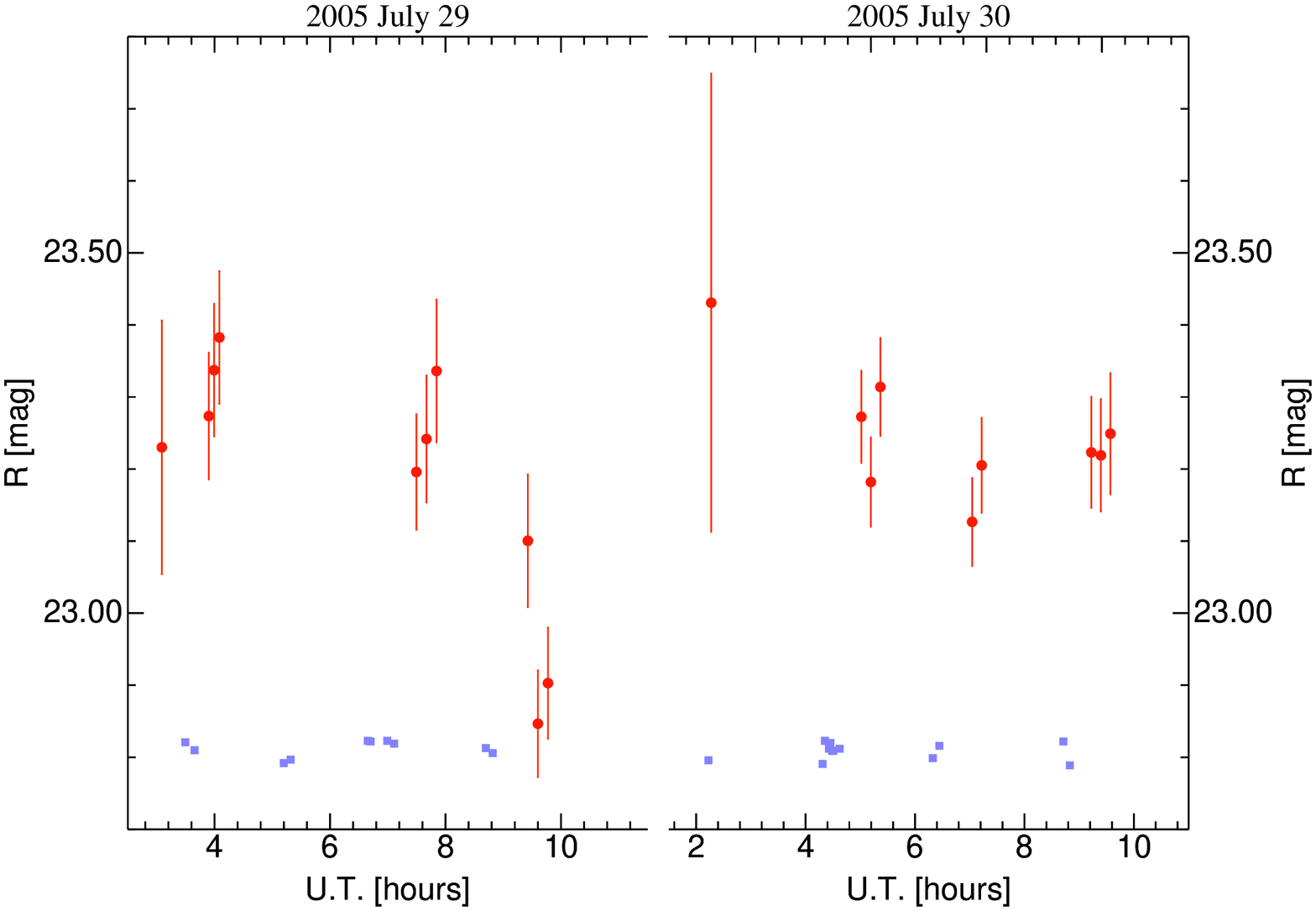}
\caption{
Light curves of \Setebos.
(See Fig.~\ref{fig1a} for comments).
  \label{fig1c}}
\end{figure*} 

\begin{figure*}
\includegraphics[angle=-90,scale=\LCSCALE]{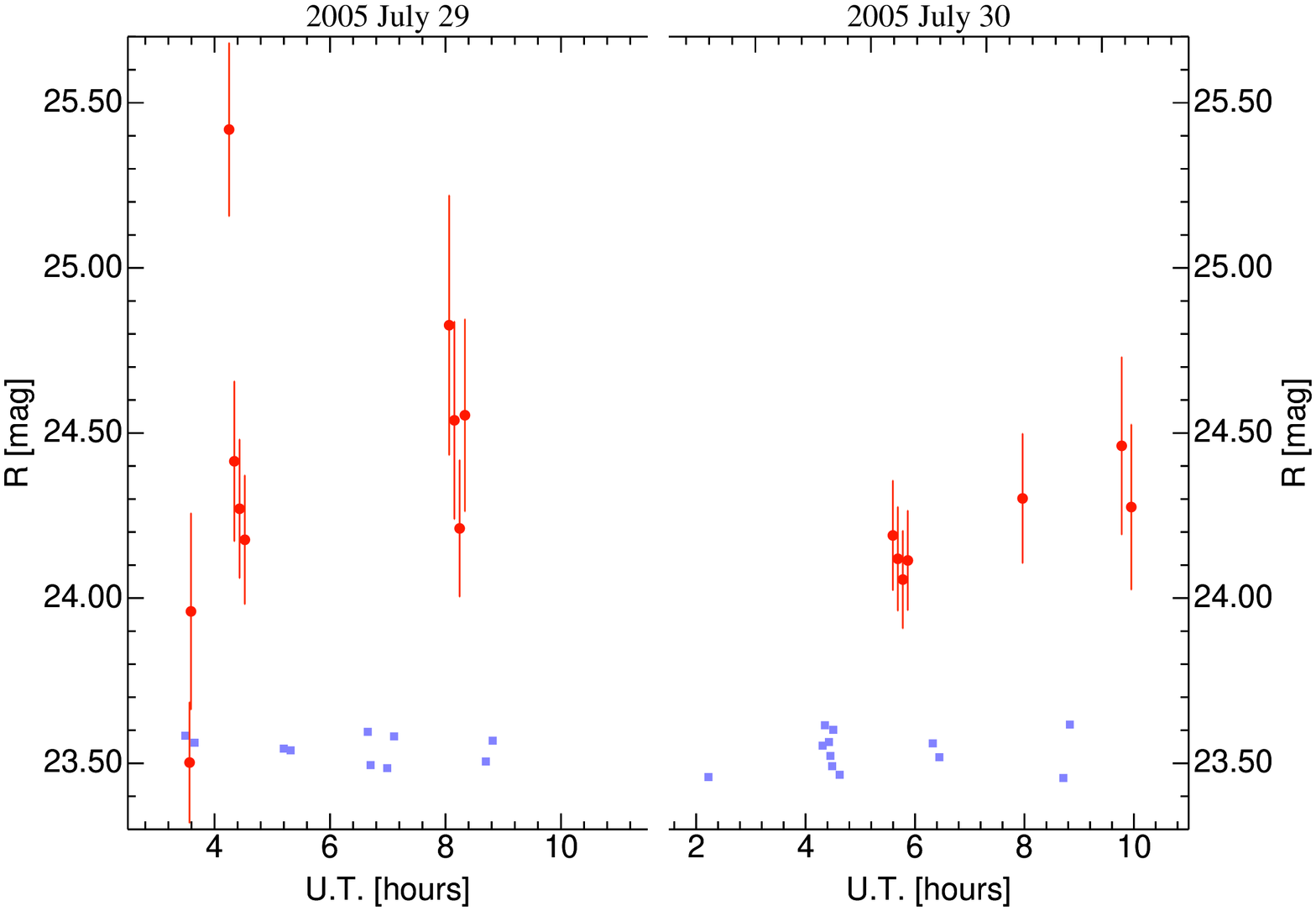}
\caption{
Light curves $R$ for \Stephano.
(See Fig.~\ref{fig1a} for comments).
  \label{fig1d}}
\end{figure*} 

\begin{figure*}
\includegraphics[angle=-90,scale=\LCSCALE]{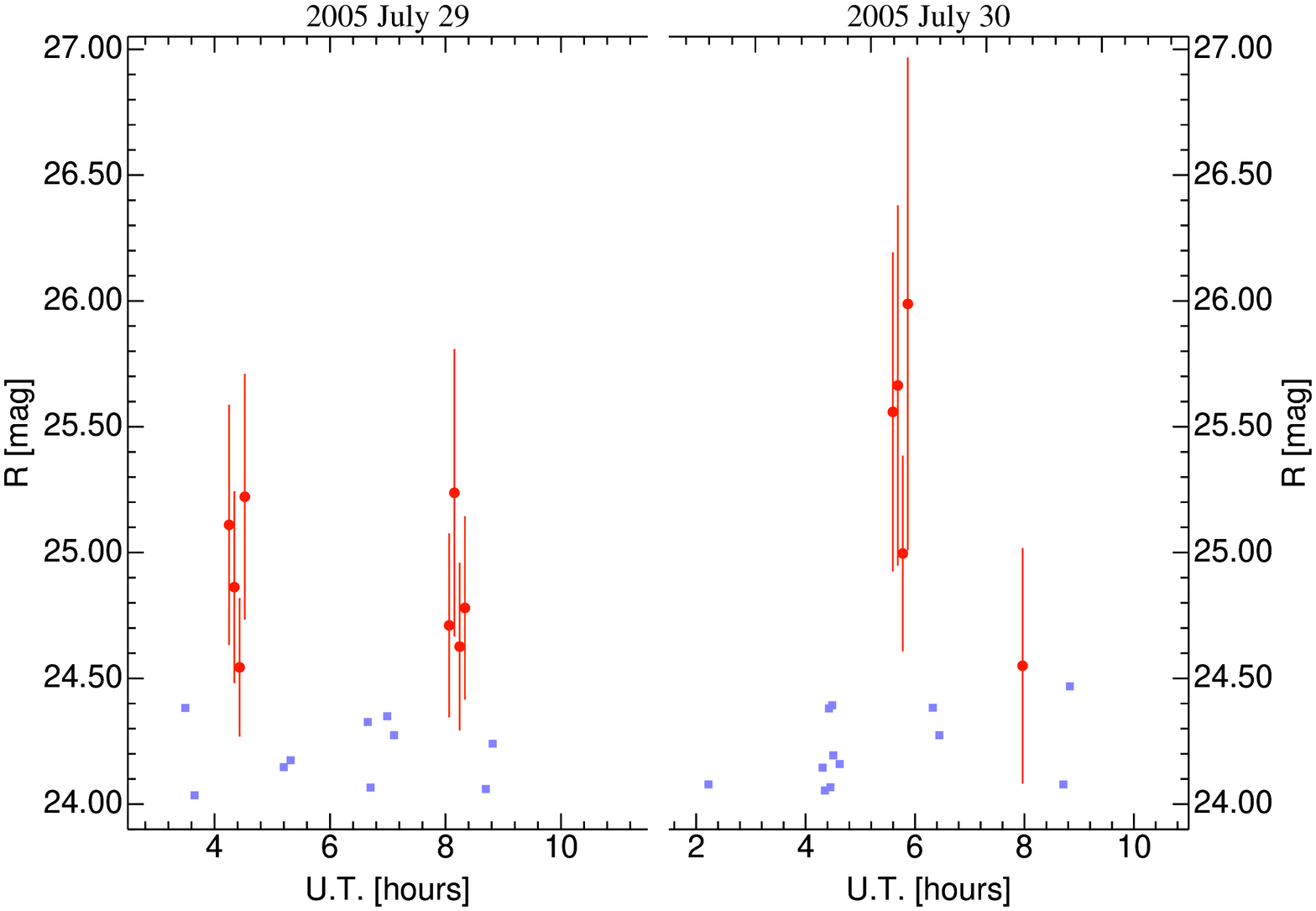}
\caption{
Light curves $R$ for \Trinculo.
(See Fig.~\ref{fig1a} for comments).
  \label{fig1e}}
\end{figure*} 

}

\newcommand{\FIGVARIABILITY}{
\begin{figure*}
\centering
\includegraphics[angle=-90, scale=0.5]{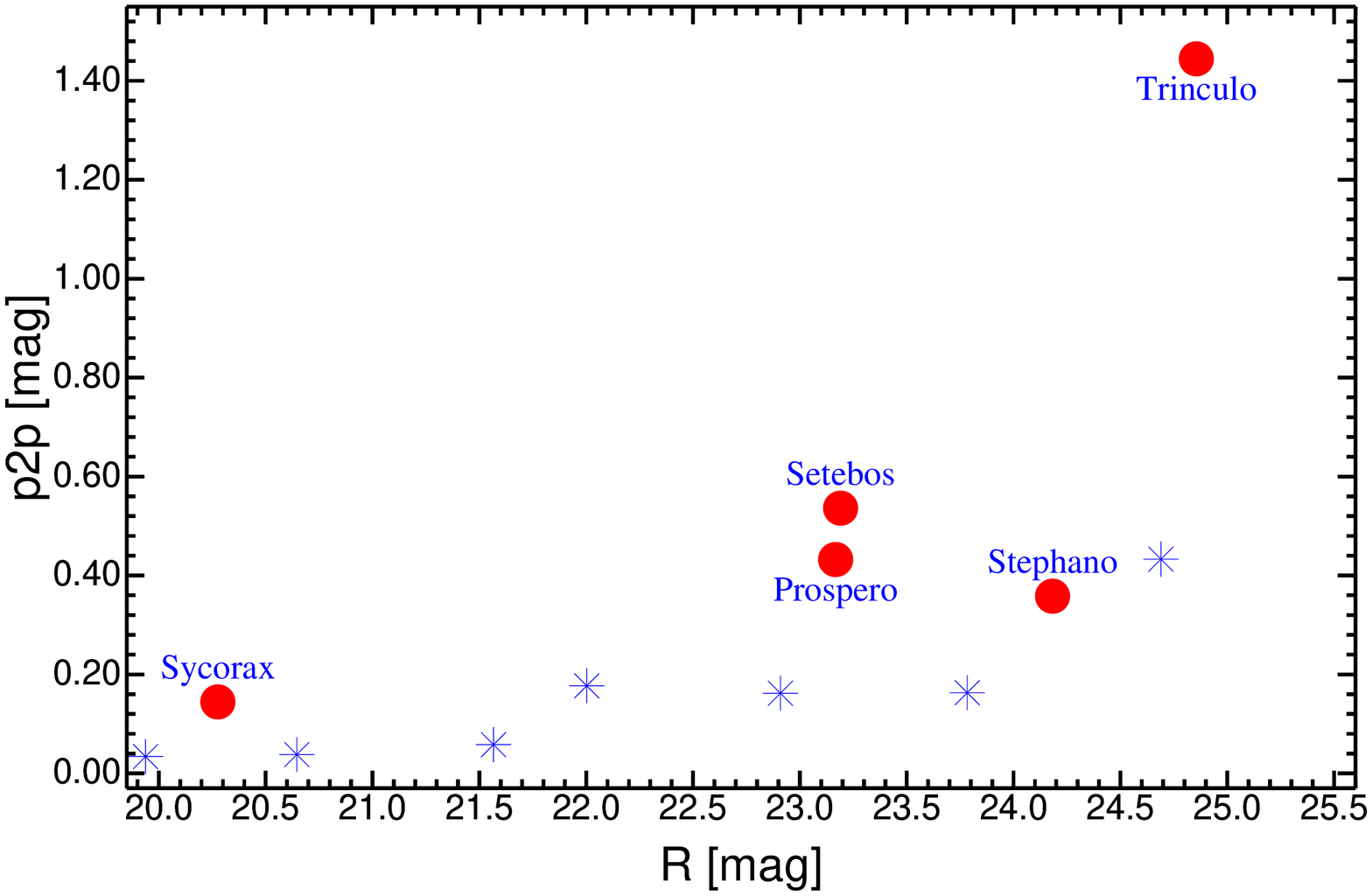} \\
\includegraphics[angle=-90, scale=0.5]{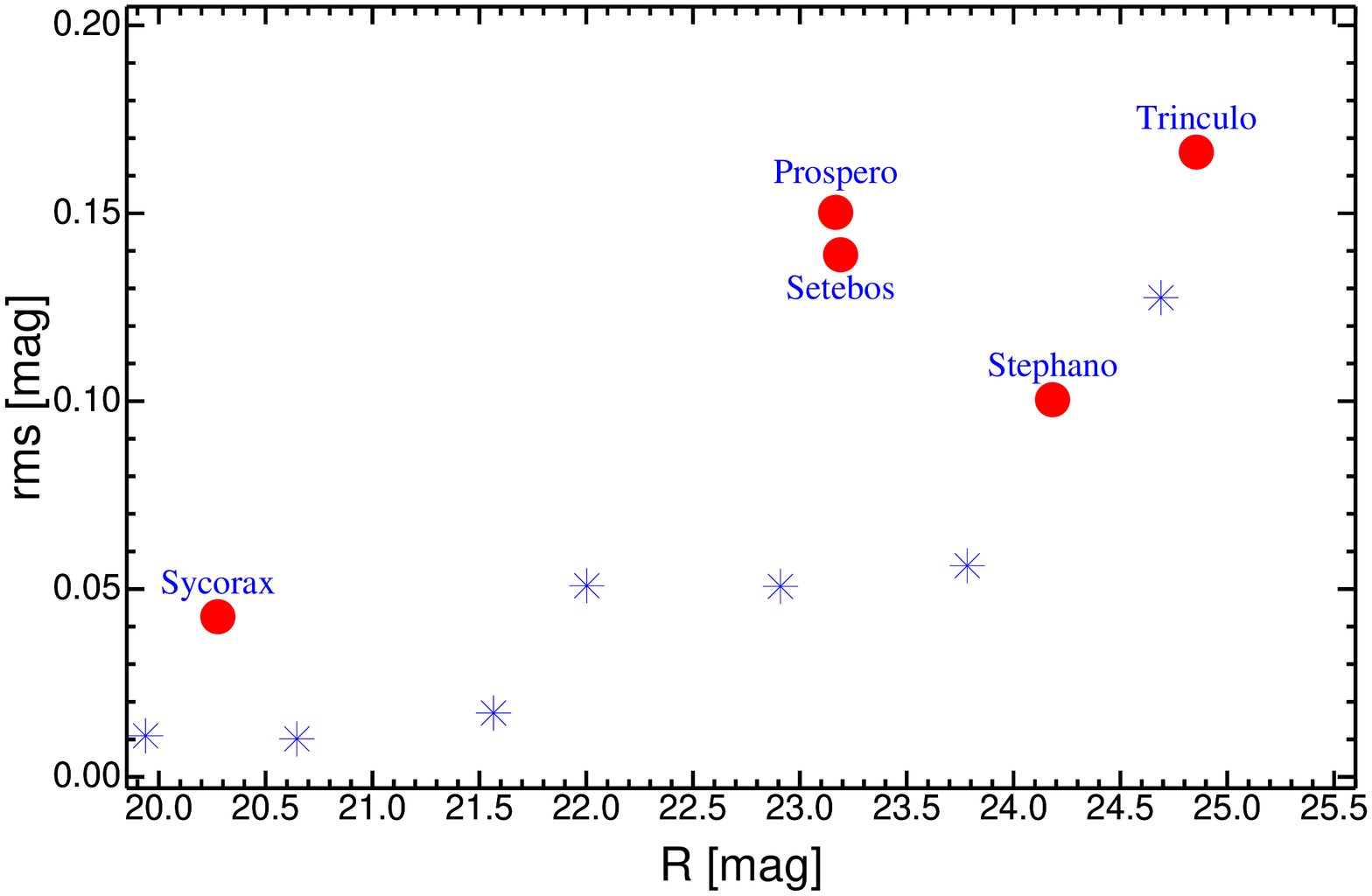}
\caption{
 Variability of the $R$ light curves for irregular satellites
(red circles) and field stars (light blue asterisks and gray band) as
a function of their magnitude. Top frame for the
peak--to--peak variation, bottom frame for RMS of the variation.
  %
  %
(See the electronic edition of the Journal for a colour version
of this figure).
  \label{fig:variability}
}
\end{figure*}
} 

\newcommand{\FIGTIMEWINDOW}{
\begin{figure*}
\centering
\vspace{1cm}
\includegraphics[scale=0.5]{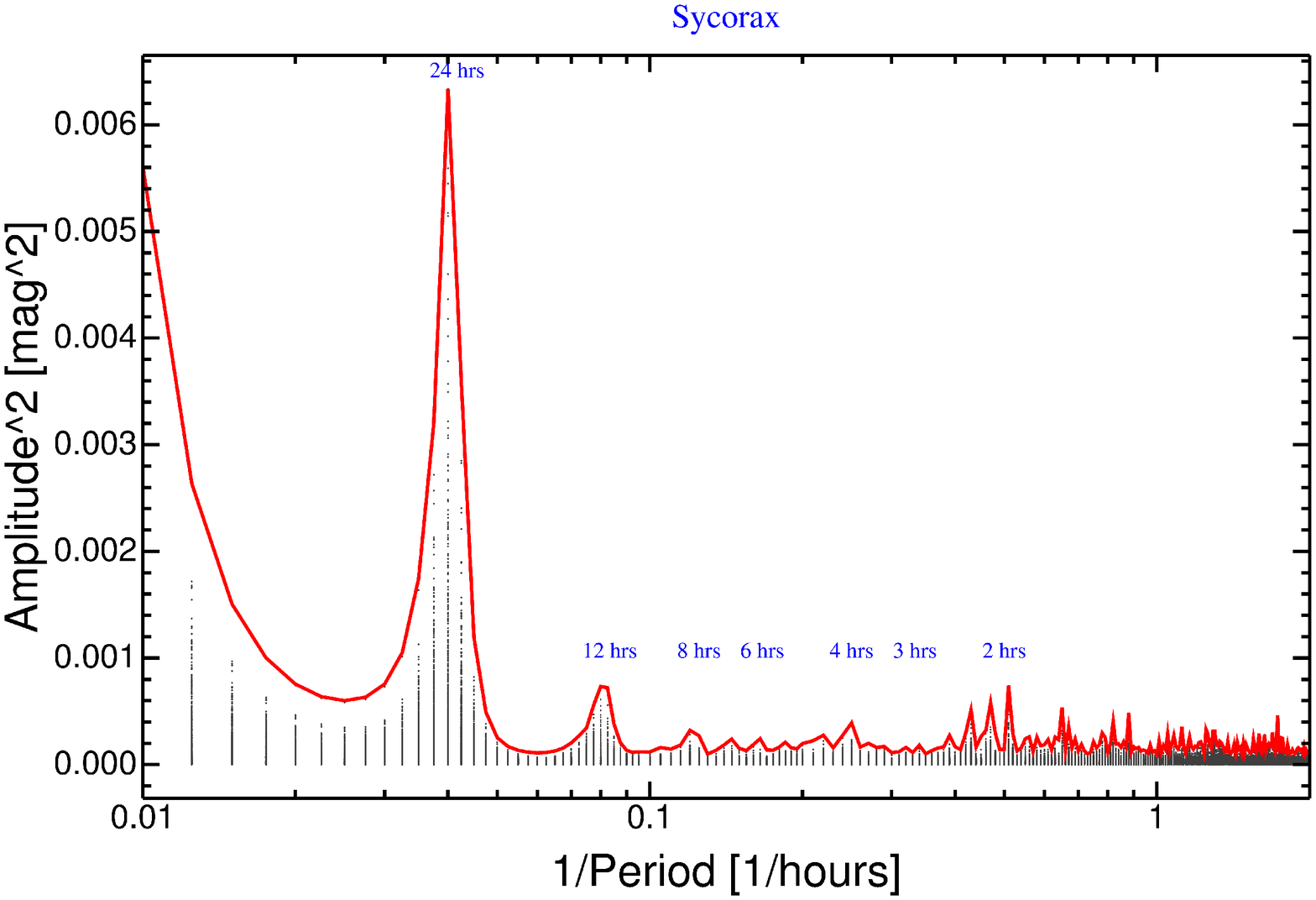} \\
\includegraphics[scale=0.5]{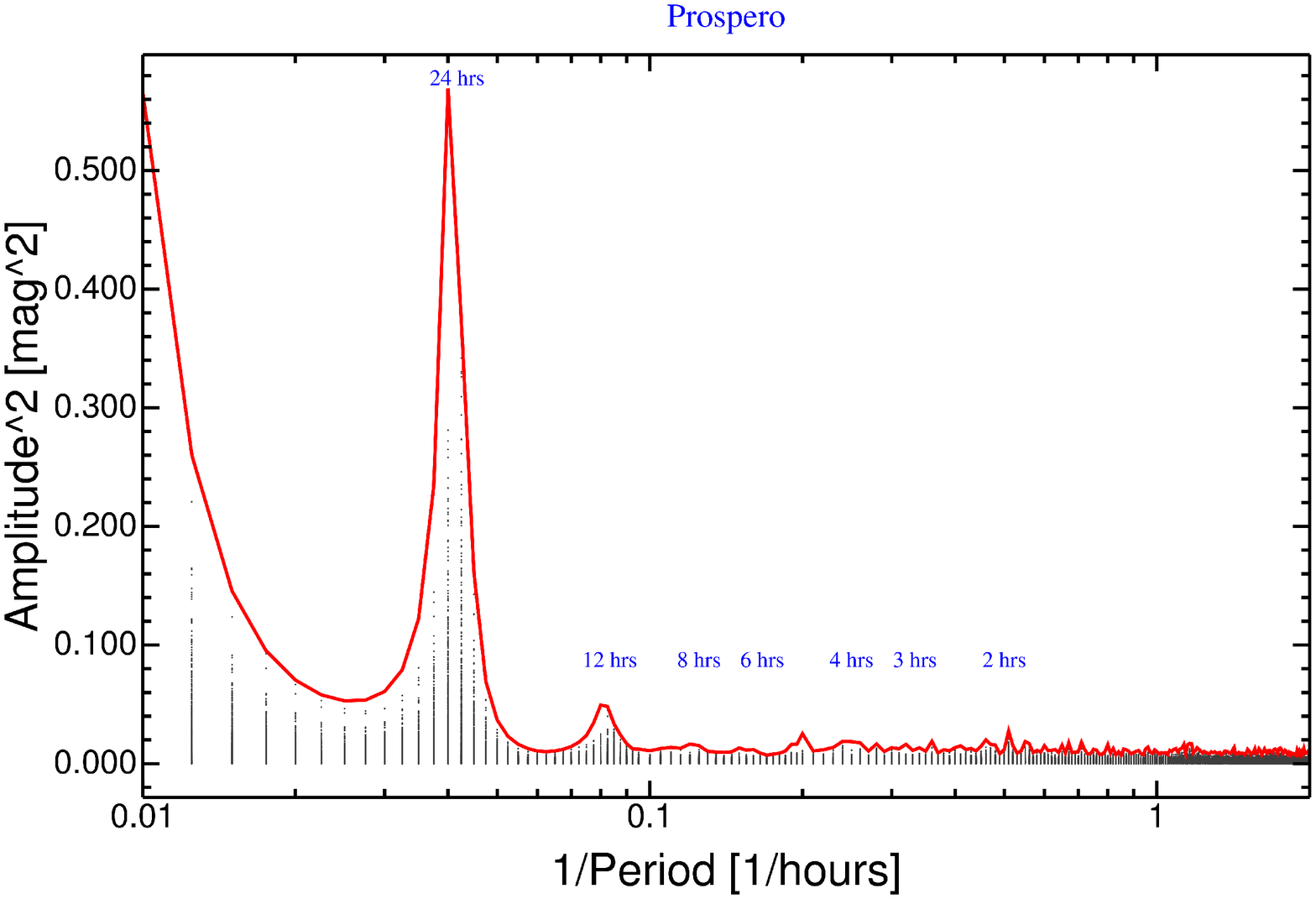}
\caption{
Time window for \Sycorax\ (upper frame) and \Prospero\  (lower frame).
(See the electronic edition of the Journal for a colour version
of this figure).
  \label{fig:time:window}
}
\end{figure*}
} 

%% file: maris_carraro_parisi_tab1.tex
\newcommand{\apex}{''}
\newcommand{\emptyA}{--- ---}
\newcommand{\Jul}{\mathrm{Jul}}
\newcommand{\Nov}{\mathrm{Nov}}
\newcommand{\emptylineA}{ & & & & & \\}

\newcommand{\tSyA}
{ 
           \Sycorax  & $    \Jul      \;      29    $ & $    03   :    27    :    13   $ &   R1  &  30  & $  20.282 \pm  0.015  $ \\ 
           \apex     & $            \apex           $ & $    04   :    40    :    37   $ &   R2  &  300 & $  20.270 \pm  0.005  $ \\ 
           \apex     & $            \apex           $ & $    04   :    46    :    05   $ &   R3  &  300 & $  20.277 \pm  0.005  $ \\ 
           \apex     & $            \apex           $ & $    06   :    46    :    49   $ &   V1  &  300 & $  20.849 \pm  0.009  $ \\ 
           \apex     & $            \apex           $ & $    06   :    52    :    17   $ &   V2  &  300 & $  20.853 \pm  0.009  $ \\ 
           \apex     & $            \apex           $ & $    08   :    29    :    17   $ &   R4  &  300 & $  20.277 \pm  0.006  $ \\ 
           \apex     & $            \apex           $ & $    08   :    34    :    45   $ &   R5  &  300 & $  20.313 \pm  0.007  $ \\ 
           \apex     & $            \apex           $ & $    09   :    12    :    19   $ &   R6  &  300 & $  20.325 \pm  0.008  $ \\ 
           \apex     & $            \apex           $ & $    09   :    17    :    47   $ &   R7  &  300 & $  20.333 \pm  0.008  $ \\ 
}
\newcommand{\tSyB}
{
           \Sycorax  & $    \Jul      \;      30    $ & $    02   :    26    :    25   $ &   R8  &  30  & $  20.214 \pm  0.023  $ \\ 
           \apex     & $            \apex           $ & $    06   :    01    :    03   $ &   R9  &  300 & $  20.283 \pm  0.006  $ \\ 
           \apex     & $            \apex           $ & $    06   :    06    :    29   $ &  R10  &  300 & $  20.289 \pm  0.006  $ \\ 
           \apex     & $            \apex           $ & $    06   :    12    :    15   $ &   B1  &  300 & $  21.701 \pm  0.013  $ \\ 
           \apex     & $            \apex           $ & $    08   :    24    :    01   $ &  R11  &  300 & $  20.215 \pm  0.006  $ \\ 
           \apex     & $            \apex           $ & $    08   :    29    :    28   $ &  R12  &  300 & $  20.203 \pm  0.006  $ \\ 
           \apex     & $            \apex           $ & $    08   :    35    :    07   $ &   V3  &  300 & $  20.739 \pm  0.008  $ \\ 
           \apex     & $            \apex           $ & $    10   :    09    :    30   $ &  R13  &  300 & $  20.322 \pm  0.009  $ \\ 
           \apex     & $            \apex           $ & $    10   :    14    :    56   $ &  R14  &  300 & $  20.348 \pm  0.013  $ \\ 
}          

\newcommand{\tStA}
{ 
           \Setebos  & $    \Jul      \;      29    $ & $    03   :    05    :    18   $ &   R1  &  60  & $  23.230 \pm  0.176  $ \\ 
           \apex     & $            \apex           $ & $    03   :    54    :    07   $ &   R2  &  300 & $  23.274 \pm  0.088  $ \\ 
           \apex     & $            \apex           $ & $    03   :    59    :    35   $ &   R3  &  300 & $  23.337 \pm  0.092  $ \\ 
           \apex     & $            \apex           $ & $    04   :    05    :    03   $ &   R4  &  300 & $  23.383 \pm  0.092  $ \\ 
           \apex     & $            \apex           $ & $    05   :    43    :    04   $ &   V1  &  600 & $  23.796 \pm  0.080  $ \\ 
           \apex     & $            \apex           $ & $    05   :    53    :    32   $ &   V2  &  600 & $  23.768 \pm  0.083  $ \\ 
           \apex     & $            \apex           $ & $    06   :    03    :    60   $ &   V3  &  600 & $  23.781 \pm  0.088  $ \\ 
           \apex     & $            \apex           $ & $    07   :    29    :    43   $ &   R5  &  600 & $  23.196 \pm  0.080  $ \\ 
           \apex     & $            \apex           $ & $    07   :    40    :    12   $ &   R6  &  600 & $  23.242 \pm  0.088  $ \\ 
           \apex     & $            \apex           $ & $    07   :    50    :    40   $ &   R7  &  600 & $  23.336 \pm  0.099  $ \\ 
           \apex     & $            \apex           $ & $    09   :    25    :    31   $ &   R8  &  600 & $  23.100 \pm  0.092  $ \\ 
           \apex     & $            \apex           $ & $    09   :    35    :    58   $ &   R9  &  600 & $  22.847 \pm  0.074  $ \\ 
           \apex     & $            \apex           $ & $    09   :    46    :    25   $ &  R10  &  600 & $  22.903 \pm  0.077  $ \\ 
}
\newcommand{\tStB}
{
           \Setebos  & $    \Jul      \;      30    $ & $    02   :    16    :    30   $ &  R11  &  60  & $  23.836 \pm  0.318  $ \\ 
           \apex     & $            \apex           $ & $    05   :    01    :    08   $ &  R12  &  600 & $  23.272 \pm  0.064  $ \\ 
           \apex     & $            \apex           $ & $    05   :    11    :    37   $ &  R13  &  600 & $  23.182 \pm  0.062  $ \\ 
           \Setebos  & $    \Jul      \;      30    $ & $    05   :    22    :    04   $ &  R14  &  600 & $  23.314 \pm  0.068  $ \\ 
           \apex     & $            \apex           $ & $    07   :    02    :    39   $ &  R15  &  600 & $  23.127 \pm  0.061  $ \\ 
           \apex     & $            \apex           $ & $    07   :    13    :    06   $ &  R16  &  600 & $  23.205 \pm  0.066  $ \\ 
           \apex     & $            \apex           $ & $    07   :    23    :    51   $ &   B1  &  600 & $  24.455 \pm  0.132  $ \\ 
           \apex     & $            \apex           $ & $    07   :    34    :    30   $ &   V4  &  600 & $  23.649 \pm  0.081  $ \\ 
           \apex     & $            \apex           $ & $    07   :    44    :    58   $ &   V5  &  600 & $  23.602 \pm  0.075  $ \\ 
           \apex     & $            \apex           $ & $    09   :    13    :    23   $ &  R17  &  600 & $  23.223 \pm  0.077  $ \\ 
           \apex     & $            \apex           $ & $    09   :    23    :    50   $ &  R18  &  600 & $  23.219 \pm  0.078  $ \\ 
           \apex     & $            \apex           $ & $    09   :    34    :    18   $ &  R19  &  600 & $  23.249 \pm  0.084  $ \\ 
}          

\newcommand{\tSpA}
{ 
           \Stephano & $    \Jul      \;      29    $ & $    03   :    34    :    06   $ &   R1  &  60  & $  23.502 \pm  0.183  $ \\ 
           \apex     & $            \apex           $ & $    03   :    35    :    34   $ &   R2  &  60  & $  23.960 \pm  0.297  $ \\ 
           \apex     & $            \apex           $ & $    04   :    15    :    10   $ &   R3  &  300 & $  25.419 \pm  0.262  $ \\ 
           \apex     & $            \apex           $ & $    04   :    20    :    37   $ &   R4  &  300 & $  24.414 \pm  0.242  $ \\ 
           \apex     & $            \apex           $ & $    04   :    26    :    04   $ &   R5  &  300 & $  24.271 \pm  0.210  $ \\ 
           \apex     & $            \apex           $ & $    04   :    31    :    31   $ &   R6  &  300 & $  24.177 \pm  0.195  $ \\ 
           \apex     & $            \apex           $ & $    06   :    16    :    34   $ &   V1  &  600 & $  25.033 \pm  0.269  $ \\ 
           \Stephano & $    \Jul      \;      29    $ & $    06   :    27    :    02   $ &   V2  &  600 & $  24.823 \pm  0.295  $ \\ 
           \apex     & $            \apex           $ & $    08   :    03    :    46   $ &   R7  &  300 & $  24.827 \pm  0.393  $ \\ 
           \apex     & $            \apex           $ & $    08   :    09    :    14   $ &   R8  &  300 & $  24.538 \pm  0.299  $ \\ 
           \apex     & $            \apex           $ & $    08   :    14    :    41   $ &   R9  &  300 & $  24.211 \pm  0.207  $ \\ 
           \apex     & $            \apex           $ & $    08   :    20    :    09   $ &  R10  &  300 & $  24.554 \pm  0.291  $ \\ 
}
\newcommand{\tSpB}
{
           \Stephano & $    \Jul      \;      30    $ & $    05   :    35    :    44   $ &  R11  &  300 & $  24.190 \pm  0.166  $ \\ 
           \apex     & $            \apex           $ & $    05   :    41    :    11   $ &  R12  &  300 & $  24.119 \pm  0.157  $ \\ 
           \apex     & $            \apex           $ & $    05   :    46    :    39   $ &  R13  &  300 & $  24.056 \pm  0.148  $ \\ 
           \apex     & $            \apex           $ & $    05   :    52    :    07   $ &  R14  &  300 & $  24.114 \pm  0.151  $ \\ 
           \apex     & $            \apex           $ & $    07   :    57    :    58   $ &  R15  &  600 & $  24.302 \pm  0.196  $ \\ 
           \apex     & $            \apex           $ & $    08   :    08    :    36   $ &   V3  &  600 & $  24.957 \pm  0.290  $ \\ 
           \apex     & $            \apex           $ & $    09   :    46    :    41   $ &  R16  &  600 & $  24.285 \pm  0.269  $ \\ 
           \apex     & $            \apex           $ & $    09   :    57    :    11   $ &  R17  &  600 & $  24.276 \pm  0.250  $ \\ 
}          

\def\tPAA{ 
           \Prospero & $    \Jul      \;      29    $ & $    03   :    29    :  44 $ &   R1  &  100 & $  23.275 \pm  0.130  $ \\ 
             \apex   & $            \apex           $ & $    03   :    39    :  16 $ &   R2  &  300 & $  23.175 \pm  0.082  $ \\ 
             \apex   & $            \apex           $ & $    03   :    46    :  44 $ &   I1  &  300 & $  22.821 \pm  0.117  $ \\ 
             \apex   & $            \apex           $ & $    05   :    12    :  00 $ &   R3  &  400 & $  23.005 \pm  0.057  $ \\ 
             \Prospero & $    \Jul      \;      29    $ & $    05   :    19    :  09 $ &   R4  &  400 & $  23.125 \pm  0.064  $ \\ 
             \apex   & $            \apex           $ & $    05   :    26    :  27 $ &   V1  &  400 & $  23.680 \pm  0.079  $ \\ 
             \apex   & $            \apex           $ & $    05   :    33    :  34 $ &   V2  &  400 & $  23.763 \pm  0.085  $ \\ 
             \apex   & $            \apex           $ & $    06   :    39    :  14 $ &   R5  &  100 & $  23.239 \pm  0.120  $ \\ 
             \apex   & $            \apex           $ & $    06   :    42    :  09 $ &   R6  &  50  & $  23.160 \pm  0.150  $ \\ 
             \apex   & $            \apex           $ & $    06   :    59    :  28 $ &   R7  &  400 & $  23.383 \pm  0.102  $ \\ 
             \apex   & $            \apex           $ & $    07   :    06    :  36 $ &   R8  &  400 & $  23.362 \pm  0.100  $ \\ 
             \apex   & $            \apex           $ & $    07   :    13    :  55 $ &   I2  &  400 & $  22.996 \pm  0.140  $ \\ 
             \apex   & $            \apex           $ & $    07   :    21    :  03 $ &   I3  &  400 & $  22.966 \pm  0.144  $ \\ 
             \apex   & $            \apex           $ & $    08   :    41    :  56 $ &   R9  &  400 & $  23.087 \pm  0.084  $ \\ 
             \apex   & $            \apex           $ & $    08   :    49    :  04 $ &  R10  &  400 & $  22.979 \pm  0.072  $ \\ 
             \apex   & $            \apex           $ & $    08   :    56    :  23 $ &   I4  &  400 & $  22.552 \pm  0.090  $ \\ 
             \apex   & $            \apex           $ & $    09   :    03    :  30 $ &   I5  &  400 & $  22.539 \pm  0.089  $ \\ 
}\def\tPAB{
           \Prospero & $    \Jul      \;      30    $ & $    02   :    13    :  32 $ &  R11  &  60  & $  23.779 \pm  0.429  $ \\ 
             \apex   & $            \apex           $ & $    04   :    18    :  45 $ &  R12  &  98  & $  22.955 \pm  0.092  $ \\ 
             \apex   & $            \apex           $ & $    04   :    21    :  15 $ &  R13  &  210 & $  23.018 \pm  0.074  $ \\ 
             \apex   & $            \apex           $ & $    04   :    25    :  37 $ &  R14  &  20  & $  23.504 \pm  0.291  $ \\ 
             \apex   & $            \apex           $ & $    04   :    27    :  11 $ &  R15  &  20  & $  22.929 \pm  0.182  $ \\ 
             \apex   & $            \apex           $ & $    04   :    29    :  12 $ &  R16  &  20  & $  22.949 \pm  0.205  $ \\ 
             \apex   & $            \apex           $ & $    04   :    30    :  18 $ &  R17  &  400 & $  23.205 \pm  0.075  $ \\ 
             \apex   & $            \apex           $ & $    04   :    37    :  26 $ &  R18  &  400 & $  23.296 \pm  0.080  $ \\ 
             \apex   & $            \apex           $ & $    04   :    44    :  43 $ &   I6  &  400 & $  22.947 \pm  0.112  $ \\ 
             \apex   & $            \apex           $ & $    04   :    51    :  51 $ &   I7  &  400 & $  22.972 \pm  0.114  $ \\ 
             \apex   & $            \apex           $ & $    06   :    19    :  32 $ &  R19  &  400 & $  23.411 \pm  0.080  $ \\ 
             \apex   & $            \apex           $ & $    06   :    26    :  39 $ &  R20  &  400 & $  23.392 \pm  0.079  $ \\ 
             \apex   & $            \apex           $ & $    06   :    34    :  20 $ &   B1  &  300 & $  24.622 \pm  0.160  $ \\ 
             \apex   & $            \apex           $ & $    06   :    39    :  48 $ &   B2  &  300 & $  24.748 \pm  0.180  $ \\ 
             \apex   & $            \apex           $ & $    06   :    45    :  40 $ &   V3  &  400 & $  23.867 \pm  0.092  $ \\ 
             \apex   & $            \apex           $ & $    06   :    52    :  47 $ &   V4  &  400 & $  23.906 \pm  0.101  $ \\ 
           \Prospero & $    \Jul      \;      30    $ & $    08   :    42    :  41 $ &  R21  &  400 & $  23.057 \pm  0.075  $ \\ 
             \apex   & $            \apex           $ & $    08   :    49    :  48 $ &  R22  &  400 & $  23.279 \pm  0.093  $ \\ 
             \apex   & $            \apex           $ & $    08   :    57    :  25 $ &   I8  &  400 & $  22.792 \pm  0.113  $ \\ 
             \apex   & $            \apex           $ & $    09   :    04    :  32 $ &   I9  &  400 & $  22.690 \pm  0.103  $ \\ 
}          

\newcommand{\tPA}{  
           \Prospero & $    \Nov      \;      22    $ & $    00   :    44    :    18   $ &   R1  &  300 & $  23.193 \pm  0.077  $ \\ 
           \apex     & $            \apex           $ & $    00   :    50    :    04   $ &   B1  &  300 & $  24.583 \pm  0.181  $ \\ 
           \apex     & $            \apex           $ & $    00   :    55    :    49   $ &   R2  &  300 & $  23.240 \pm  0.081  $ \\ 
           \apex     & $            \apex           $ & $    01   :    01    :    27   $ &   V1  &  300 & $  23.881 \pm  0.110  $ \\ 
           \apex     & $            \apex           $ & $    01   :    07    :    06   $ &   R3  &  300 & $  23.310 \pm  0.090  $ \\ 
           \apex     & $            \apex           $ & $    01   :    12    :    45   $ &   I1  &  300 & $  22.726 \pm  0.157  $ \\ 
           \apex     & $            \apex           $ & $    01   :    15    :    43   $ &   I2  &  300 & $  22.925 \pm  0.243  $ \\ 
           \apex     & $            \apex           $ & $    01   :    18    :    52   $ &   R4  &  300 & $  23.120 \pm  0.076  $ \\ 
           \apex     & $            \apex           $ & $    01   :    24    :    38   $ &   B2  &  300 & $  24.710 \pm  0.218  $ \\ 
           \apex     & $            \apex           $ & $    01   :    30    :    25   $ &   R5  &  300 & $  23.212 \pm  0.083  $ \\ 
}
\newcommand{\tPB}
{ 
           \Prospero & $    \Nov      \;      25    $ & $    00   :    36    :    16   $ &   V2  &  300 & $  23.835 \pm  0.109  $ \\ 
           \apex     & $            \apex           $ & $    00   :    41    :    54   $ &   R6  &  300 & $  23.194 \pm  0.080  $ \\ 
           \apex     & $            \apex           $ & $    00   :    47    :    32   $ &   I3  &  300 & $  22.838 \pm  0.184  $ \\ 
           \apex     & $            \apex           $ & $    00   :    50    :    29   $ &   I4  &  300 & $  22.969 \pm  0.207  $ \\ 
           \apex     & $            \apex           $ & $    00   :    53    :    37   $ &   R7  &  300 & $  23.279 \pm  0.088  $ \\ 
           \apex     & $            \apex           $ & $    00   :    59    :    23   $ &   B3  &  300 & $  24.686 \pm  0.203  $ \\ 
           \apex     & $            \apex           $ & $    01   :    05    :    07   $ &   R8  &  300 & $  23.063 \pm  0.076  $ \\ 
           \apex     & $            \apex           $ & $    01   :    10    :    45   $ &   V3  &  300 & $  23.837 \pm  0.108  $ \\ 
           \apex     & $            \apex           $ & $    01   :    16    :    22   $ &   R9  &  300 & $  23.284 \pm  0.165  $ \\ 
}          

\newcommand{\tTyA}
{ 
           \Trinculo & $    \Jul      \;      29    $ & $    04   :    15    :    10   $ &   R3  &  300 & $  25.110 \pm  0.474  $ \\ 
           \apex     & $            \apex           $ & $    04   :    20    :    37   $ &   R4  &  300 & $  24.862 \pm  0.378  $ \\ 
           \apex     & $            \apex           $ & $    04   :    26    :    04   $ &   R5  &  300 & $  24.544 \pm  0.272  $ \\ 
           \apex     & $            \apex           $ & $    04   :    31    :    31   $ &   R6  &  300 & $  25.222 \pm  0.485  $ \\ 
           \apex     & $            \apex           $ & $    06   :    16    :    34   $ &   V1  &  600 & $  25.925 \pm  0.671  $ \\ 
           \apex     & $            \apex           $ & $    06   :    27    :    02   $ &   V2  &  600 & $  26.189 \pm  0.901  $ \\ 
           \apex     & $            \apex           $ & $    08   :    03    :    46   $ &   R7  &  300 & $  24.711 \pm  0.362  $ \\ 
           \apex     & $            \apex           $ & $    08   :    09    :    14   $ &   R8  &  300 & $  25.237 \pm  0.568  $ \\ 
           \apex     & $            \apex           $ & $    08   :    14    :    41   $ &   R9  &  300 & $  24.626 \pm  0.330  $ \\ 
           \apex     & $            \apex           $ & $    08   :    20    :    09   $ &  R10  &  300 & $  24.780 \pm  0.361  $ \\ 
}
\newcommand{\tTyB}
{
           \Trinculo & $    \Jul      \;      30    $ & $    05   :    35    :    44   $ &  R11  &  300 & $  25.559 \pm  0.631  $ \\ 
           \apex     & $            \apex           $ & $    05   :    41    :    11   $ &  R12  &  300 & $  25.664 \pm  0.713  $ \\ 
           \apex     & $            \apex           $ & $    05   :    46    :    39   $ &  R13  &  300 & $  24.996 \pm  0.386  $ \\ 
           \apex     & $            \apex           $ & $    05   :    52    :    07   $ &  R14  &  300 & $  25.988 \pm  0.976  $ \\ 
           \apex     & $            \apex           $ & $    07   :    57    :    58   $ &  R15  &  600 & $  24.550 \pm  0.465  $ \\ 
           \apex     & $            \apex           $ & $    08   :    08    :    36   $ &   V3  &  600 & $  25.167 \pm  0.646  $ \\ 
}          

\longtab{1}{
 \begin{longtable}{cccccc}
  \caption{Log of observations. 
  U.T. refers to the start times of exposures and are not corrected for
  light travel time. The Flt. column referes to the filter (BVRI)
  and frame number obtained 
  with that filter (eg.: R3 is the third R frame for the given object in 
  the serie).    $\Texp$ is the exposure time in seconds, the shortest 
  exposures have been acquired to improve frame centering. 
   } \label{tab:mag}\\
  \hline \hline
  Obj &
  Epoch &
  U.T. &
  Flt. &
  {$\stackrel{\Texp}{[sec]}$} &
  mag \\
  \hline
 \endfirsthead
  \caption{Continued.}\\
  \hline\hline
  Obj &
  Epoch &
  U.T. &
  Flt. &
  {$\stackrel{\Texp}{[sec]}$} &
  mag \\
   \hline
 \endhead
  \hline
 \endfoot
\tSyA
\tSyB
\tSpA
\tSpB
\tStA
\tStB
\tTyA
\tTyB
\tPAA
\tPAB
\tPA
\tPB
\end{longtable}
} 

%% file: maris_carraro_parisi_tab2.tex
 \setcounter{table}{1}
 \begin{table}[t]
 \centering
 \caption{Testing against random fluctuations.
 Columns  2 - 3 $\chi^2$ and significativity level (SL) 
assuming constant signal,
         4 - 5 $\chi^2$ and SL assuming linear time dependence,
         6 - 7 maximum amplitude and false allarm probability
assuming periodical signal.
Note that in all the cases a low probability (either
SL or \PFA) denotes
a high level of confidence in excluding noise fluctuations.
\label{tab:random:tests}
 }
 \begin{tabular}{lcccccc}
 \hline\hline
           & \multicolumn{6}{c}{Hypothesis H0}\\
           & \multicolumn{2}{c}{Constant} &
             \multicolumn{2}{c}{Linear Trend} &
             \multicolumn{2}{c}{Periodical} \\
 Object    &
             $\chi^2_{\mathrm{const}}$ & SL                &
             $\chi^2_{\mathrm{linear}}$ & SL                &
             $\Amax$     & \PFA \\
 \hline
 \Sycorax &
       557    & $<1\times10^{-9}$   &
       437    & $<1\times10^{-9}$   &
        0.07  & $<1\times10^{-8}$
\\
 \Prospero &
      92.7     & $1\times10^{-7}$   &
      85.2     & $1\times10^{-6}$   &
      0.22  &   $<3\times10^{-6}$
\\
 %
%
\Stephano &
    45.3 &   $4\times10^{-4}$ &
    44.9 &   $3\times10^{-4}$ &
    0.36 &   0.22
\\
\Setebos &
    59.2 & $3\times10^{-5}$ &
    59.2 & $2\times10^{-5}$ &
    0.19 & $7\times10^{-3}$
\\
\Trinculo &
    8.83 & $8\times10^{-1}$  &
    8.44 & $8\times10^{-1}$  &
    0.42 & 0.44
\\
 \hline\hline
 \end{tabular}
 \end{table}

%% file: maris_carraro_parisi_tab3.tex
 \begin{table}[t]
 \centering
 \caption{
 Possible periods and amplitudes from fitting of
 Eq.~\ref{eq:sin:model}. 
 Column 1 $\chi^2$ for fitting, column 2 the confidence level (CL), 
 Column 3 the best fit period in hours and its estimated internal
 uncertainty, Column 4 the corresponding amplitudes and their
 uncertainties.
Solutions are ordered with increasing
 $\chi^2$. 
 \label{tab:sin:fit} 
 }
 \begin{tabular}{ccccc}
 \hline\hline
   \multicolumn{3}{c}{\bf \Sycorax}      & $P$  & $A$   \\
 \# & $\chi^2$ & CL & $[$hr$]$        & $[$mag$]$ \\
 1  & 90.734   &  -- & $3.60 \pm 0.02$ & $0.067 \pm 0.004$ \\
 2  & 134.88   &  $2\sigma$ & $2.70 \pm 0.03$ & $0.065 \pm 0.003$ \\
 3  & 179.59   &  $3\sigma$ & $3.04 \pm 0.02$ & $0.051 \pm 0.010$ \\
 4  & 180.66   &  $3\sigma$ & $3.13 \pm 0.02$ & $0.051 \pm 0.010$ \\
 \hline
    &          &    &      &       \\
   \multicolumn{3}{c}{\bf \Prospero}      & $P$  & $A$   \\
 \# & $\chi^2$ & CL & $[$hr$]$ & $[$mag$]$ \\
 \hline
 1  & 27.288 & --         & $4.551 \pm 0.040$ & $0.221 \pm 0.027$ \\
 2  & 38.759 & $2\sigma$ & $3.827 \pm 0.064$ & $0.201 \pm 0.029$ \\
 3  & 53.286 & $3\sigma$ & $5.760 \pm 0.100$ & $0.162 \pm 0.090$ \\
 4  & 55.079 & $3\sigma$ & $3.300 \pm 0.100$ & $0.121 \pm 0.090$ \\
 \hline\hline
 \end{tabular}
 \end{table}

%% file: maris_carraro_parisi_tab4.tex
 \begin{table*}
 \centering
 \caption{Weighted averages of magnitudes \label{tab:aver:mag}}
 \begin{tabular}{lccccc}
 \hline \hline
  {Obj} &
  {Run} &
  {B} &
  {V} &
  {R} &
  {I}
\\
 \hline 
\Sycorax  & Jul. & $21.701 \pm 0.013$ & $20.807 \pm 0.005$ & $20.276 \pm 0.002$ & \empty \\
\Prospero & Jul. & $24.678 \pm 0.120$ & $23.788 \pm 0.044$ & $23.160 \pm 0.019$ & $22.760 \pm 0.036$\\
\Prospero & Nov. & $ 24.651 \pm 0.115$ & $23.851 \pm 0.063$& $23.196 \pm 0.028$ &$22.864 \pm 0.095$\\
\Setebos  & Jul. & $24.455 \pm 0.132$ & $23.713 \pm 0.036$ & $23.192 \pm 0.018$ & \empty \\
\Stephano & Jul. & \empty & $24.944 \pm 0.164$ & $24.212 \pm 0.050$ & \empty \\
\Trinculo &  Jul. & \empty & $25.670 \pm 0.413$ & $24.855 \pm 0.117$ & \empty \\
 \hline \hline
\end{tabular}
\end{table*}

%% file: maris_carraro_parisi_tab6.tex
 \begin{table}
 \centering
 \caption{Colours derived from the hierarchical method\label{tab:itp:col}}
 \begin{tabular}{lcccc}
 \hline \hline
  {Obj} &
  {Run} &
  {B-V} &
  {V-R} &
  {R-I}
  \\
 \hline
\Sycorax  & Jul. & $0.915 \pm 0.017$ & $0.548 \pm 0.006$ & \\
\Prospero & Jul. & $0.793 \pm 0.138$ & $0.590 \pm 0.056$ & $0.395 \pm 0.046$\\
\Prospero & Nov. & $0.792 \pm 0.171$ & $0.635 \pm 0.085$ & $0.335 \pm 0.131$ \\
\Setebos  & Jul  & $0.831 \pm 0.143$ & $0.444 \pm 0.053$ & \\
 \hline \hline
\end{tabular}
\end{table}

%% file: maris_carraro_parisi_tab5.tex
 \begin{table}
 \centering
 \caption{Colours derived from weigthed averages \label{tab:aver:col}}
 \begin{tabular}{lcccc}
 \hline \hline
  {Obj} &
  {Run} &
  {B-V} &
  {V-R} &
  {R-I}
  \\
 \hline
\Sycorax  & Jul. & $0.893 \pm 0.014$ & $0.531 \pm 0.005$ & \empty \\
\Prospero & Jul. & $0.890 \pm 0.127$ & $0.628 \pm 0.048$ & $0.400 \pm 0.041$\\
\Prospero & Nov. & $0.800 \pm 0.131$ & $0.655 \pm 0.069$ & $0.332 \pm 0.099$\\
\Setebos  & Jul. & $0.741 \pm 0.137$ & $0.522 \pm 0.041$ & \empty \\
\Stephano & Jul. & \empty & $0.732 \pm 0.171$ &  \empty \\
\Trinculo & Jul. & \empty & $0.815 \pm 0.430$ & \empty \\
 \hline \hline
\end{tabular}
\end{table}